\documentclass{aastex61}

\newcommand{\ad}{\nabla_{\rm ad}}
\newcommand{\fiso}{F_{\rm iso}}
\newcommand{\fgas}{F_{\rm gas}}
\newcommand{\hpeb}{H_{\rm peb}}
\newcommand{\hgas}{H_{\rm gas}}
\newcommand{\matmos}{M_{\rm atmos}}
\newcommand{\mcore}{M_{\rm core}}
\newcommand{\mearth}{M_\oplus}
\newcommand{\mdisk}{M_{\rm disk}}
\newcommand{\meff}{M_{\rm mig\ eff}}
\newcommand{\mgap}{M_{\rm gap}}
\newcommand{\minv}{M_{\rm inv}}
\newcommand{\miso}{M_{\rm iso}}
\newcommand{\mjup}{M_{\rm Jup}}
\newcommand{\mplan}{M_{\rm planet}}
\newcommand{\mstar}{M_\ast}
\newcommand{\mvis}{M_{\rm vis}}
\newcommand{\niter}{N_{\rm iter}}
\newcommand{\npart}{N_{\rm part}}
\newcommand{\nsys}{N_{\rm sys}}

\newcommand{\rcap}{R_{\rm cap}}
\newcommand{\rearth}{R_\oplus}

\newcommand{\rinner}{r_{\rm inner}}
\newcommand{\router}{r_{\rm outer}}
\newcommand{\sigmagas}{\Sigma_{\rm gas}}
\newcommand{\sigmapeb}{\Sigma_{\rm peb}}
\newcommand{\stefan}{\sigma_{\rm sb}}
\newcommand{\st}{\rm St}
\newcommand{\tcool}{t_{\rm cool}}
\newcommand{\tdisk}{t_{\rm disk}}
\newcommand{\tpeb}{t_{\rm peb}}

\newcommand{\vfrag}{v_{\rm frag}}

\newcommand{\vkep}{v_{\rm kep}}
\newcommand{\vrel}{v_{\rm rel}}

\received{}
\revised{}
\accepted{\today}
\submitjournal{ApJ}

\shorttitle{Optimized Population Synthesis}
\shortauthors{Chambers}

\begin{document}

\title{Planet Formation: An Optimized Population-Synthesis Approach}

\correspondingauthor{John Chambers}
\email{jchambers@carnegiescience.edu}

\author{John Chambers}
\affil{Carnegie Institution for Science \\
Department of Terrestrial Magnetism, \\
5241 Broad Branch Road, NW, \\
Washington, DC 20015, USA}

%
%
\begin{abstract}
The physics of planet formation is investigated using a population synthesis approach. We develop a simple model for planetary growth including pebble and gas accretion, and orbital migration in an evolving protoplanetary disk. The model is run for a population of 2000 stars with a range of disk masses and radii, and initial protoplanet orbits. The resulting planetary distribution is compared with the observed population of extrasolar planets, and the model parameters are improved iteratively using a particle swarm optimization scheme. The characteristics of the final planetary systems are mainly controlled by the pebble isolation mass, which is the mass of a planet that perturbs nearby gas enough to halt the inward flux of drifting pebbles and stop growth. The pebble isolation mass increases with orbital distance so that giant planet cores can only form in the outer disk. Giants migrate inwards, populating a wide range of final orbital distances. The best model fits have large initial protoplanet masses, short pebble drift timescales, low disk viscosities, and short atmospheric cooling times, all of which promote rapid growth. The model successfully reproduces the observed frequency and distribution of giant planets and brown dwarfs. The fit for super Earths is poorer for single-planet systems, but improves steadily when more protoplanets are included. Although the study was designed to match the extrasolar planet distribution, analogs of the Solar System form in 1--2\% of systems that contain at least 4 protoplanets.

\end{abstract}


%
%
\section{Introduction} \label{sec:intro}
Planetary systems exhibit a remarkable degree of diversity in terms of the number of observed planets per system, the masses of these planets, their densities, and their orbits \citep{winn:2015}. Presumably, all these systems formed as a result of a common set of physical processes acting on protoplanetary disks that were qualitatively similar, at least initially. Thus, the observed diversity demands an explanation. Differences between planetary systems could have arisen for several reasons including differences in the initial disks, different environmental conditions, or chance events during planet formation. The relative importance of these factors is unclear, however.

Numerical models for the formation of planetary systems can help address this question, and may tell us why a particular system looks the way it does. Unfortunately, our current understanding of planet formation is far from complete, and this limits the usefulness of these models. For example, the main physical process driving the evolution of gaseous protoplanetary disks remains uncertain. Conventional ``$\alpha$-disk'' models are difficult to reconcile with observations \citep{rafikov:2017}. Recent work has focussed on several evolution mechanisms including disk winds, hydrodynamic turbulence, and planetary torques \citep{suzuki:2016, bai:2016, hartmann:2018, fung:2017}, although the relative importance of these remains unclear. As a result, the dynamics of small, solid particles is also somewhat unclear. This in turn affects our picture of the growth, radial mixing, and histories of dust grains and pebble-size particles \citep{ciesla:2007, birnstiel:2010, bitsch:2018}. 

Another major area of uncertainty is the mechanism by which pebbles are converted into asteroid-size planetesimals, as well as the time scale and efficiency of this conversion \citep{cuzzi:2010, johansen:2011, birnstiel:2016}. We cannot say with certainty whether the main building blocks of solid planets and the cores of gas-giant planets are pebbles, planetesimals, or both, nor precisely how the building blocks change with time and location in a disk. Finally, it is apparent that planetary orbits can evolve during and after planet formation \citep{ward:1997, trilling:1998}. However, the range of circumstances in which orbital migration is important is still being investigated, and the extent of migration in a typical planetary system is unclear.

Population synthesis models provide one approach to reducing these uncertainties. These studies adopt relatively simple models for disk evolution, planetary growth, and planetary migration. The main uncertainties in the physics of planet formation are encapsulated in model parameters. Typically, models are run many times, and the results compared with the Solar System or the observed extrasolar planet distribution to constrain the parameter values.

\citet{ida:2004} conducted one of the earliest population synthesis studies using an analytic model for the growth of planets accreting planetesimals and gas in a weakly viscous disk. The simulations included ``type-II'' migration in which the orbits of planets massive enough to open a gap migrate with the viscous evolution of the disk \citep{ward:1997}. The authors followed the evolution of a large number of single-planet systems, with planets starting at random locations in disks with a range of disk masses. The simulations produced populations of terrestrial and gas-giant planets somewhat similar to the observed distribution with a dearth of planets at intermediate masses inside a few AU. At larger distances, planets of all masses formed. The study also produced a large population of giant planets orbiting close to the star, which is not seen by observational surveys. We note that the model used by \citet{ida:2004} was essentially deterministic, so that the variety of final planetary systems arose due to different initial conditions.

In a follow-up study, \citet{ida:2008} added ``type-I'' migration in which planets that are too small to open a gap in the disk migrate due to linear tidal torques \citep{ward:1997}. These simulations yielded a population of small planets at a range of distances, and a pile up of planets at the inner edge of the disk, in common with the earlier study. However, planets always migrated to the inner edge of the disk before they could become gas giants unless type-I migration rates were artificially reduced by at least an order of magnitude. With reduced migration rates, planets that grew at early times were often lost, but planets growing at later times could survive, and some became gas giants.

\citet{mordasini:2009a,mordasini:2009b} used a somewhat more detailed semi-analytic model to study the growth of single-planet systems using a range of disk conditions. Both type-I and type-II migration were included. The simulations yielded a diverse population of planets including giants, provided that type-I migration rates were reduced by 1--3 orders of magnitude. The authors also determined which subset of their sample (mainly giant planets) would be observable using existing Doppler radial velocity surveys. Comparing the mass and orbital distributions of this subset with the observed population, the authors found an acceptable match, again provided that type-I rates were greatly reduced.

\citet{ida:2010} carried out a study using multiple planets per system. Dynamical interactions between planets were neglected apart from the possibility that planets undergoing converging migration could be trapped in mean-motion resonances. However, planets in the same system competed for resources, so that the growth of one planet was affected by the presence of its neighbors. The authors ignored gas accretion, focussing instead on the growth and evolution of solid planets. They found that small planets in the inner disk commonly captured one another in resonances, subsequently migrating together through the disk. The end result was a population of low-mass planets on short-period orbits qualitatively similar to the population of ``super-Earths'' found by the Kepler mission.

\citet{alibert:2013} also considered multiple planets per system, but included dynamical interactions between the largest planets using an $N$-body integrator. Thus, their model added the effects of chance events during the evolution due to the chaotic orbital interactions between planets. The study compared the outcome of ensembles of simulations using single-planet and 10-planet systems. The authors found that  the population of giant planets produced in each case was quite similar, suggesting that the formation of giant planets is not greatly affected by neighbors or chance events. However, the population of low-mass planets in each case was very different. The simulations with multiple planets generated a large population of short-period Earth and sub-Earth mass planets that was absent from the single planet case. 

\citet{ndugu:2018} looked at planet formation in disks subject to heating from other stars in their birth cluster,  thus adding an element of environmental diversity to planet formation. The authors carried out a population synthesis of single-planet systems for planets accreting pebbles and gas, and subject to migration. The study found that planet formation via pebble accretion is sensitive to the local cluster temperature, primarily because it affects the disk scale height and pebble accretion efficiency in the outer disk. The authors concluded that the efficiency of giant planet formation, and its dependence on stellar metallicity, could be strongly affected by disk heating due to nearby stars.

Thanks to observational surveys, we now have a large dataset of extrasolar systems that can be used as tests for population synthesis studies of planet formation. These surveys have found a broad variety of systems, most of which are very different than the Solar System. Planets with periods less than 100 days and masses between that of Earth and Neptune seem to be very common \citep{fulton:2017}, although there is no analog in the Solar System. Giant planets are relatively uncommon by comparison \citep{cumming:2008}.

However, the observed planetary distribution is subject to a number of observational biases. For example, the Doppler radial velocity method tends to find planets that impart large reflex motions on their star, which favors the discovery of massive, short-period planets. Most discoveries are announced after a planet has completed at least one orbital period, which limits the maximum period to values comparable to the length of the survey. The transit method is even more biased towards discovering short-period planets, and tends to find planets with large radii, which favors planets with large mass, low density or both.

Various studies have tried to determine the underlying planetary distribution taking into account observational biases. These studies often estimate the fraction of stars with planets in a certain range of orbital periods and a range of masses or radii. Some other studies estimate the number of planets per star with certain planetary characteristics. We will make use of several estimates of each kind here.

In this paper, we adopt a population synthesis approach to examine the physics of planet formation. In common with many earlier studies, we consider the possibility that much of the observed diversity of planetary systems is due to differences in initial conditions. Unlike most existing studies, we focus on the growth of planets by accreting pebbles  rather than planetesimals, since recent work suggests pebble accretion plays an important role in planet formation, especially the formation of giant planets \citep{lambrechts:2012}.

Like some earlier works, we quantify how well the synthetic planetary distribution produced by our model matches observations, taking into account observational biases.  However, we go further by iterating this procedure using an optimization scheme to successively refine the parameter values for the model. In this way we can determine the best parameter values that match the available observations, and thus help to constrain the physics of planet formation.

The rest of the paper is organized as follows. Section~2 describes the planet formation model we use.  Section~3 gives details of the observational constraints used to test the model. Section~4 describes how the model parameters are optimized iteratively by comparing the model output with observations. In Section~5 we look at the results for single planet systems, while Section~6 describes the results for multi-planet systems. Section~7 examines Solar System analogs formed by the best-fit models. Section~8 contains a discussion, and the main results are summarized in Section~9.

%
%
\section{Planet Formation Model} \label{sec:model}
For this study we require a planet-formation model that has several characteristics. Firstly, it should be very efficient so that it can be run for many iterations of the optimization scheme with different parameters, running for a large number of planetary systems in each case. Secondly, the model should include the most important features of planet formation: the growth of a solid planet, gas accretion, and orbital migration, all taking place in an evolving protoplanetary disk. Thirdly, the model should contain a small number of parameters that encapsulate the main uncertainties in the physics of planet formation.

While the model we use is quite general, we do make some assumptions. In particular, we consider the growth of a small number of protoplanets that form early in a disk's lifetime, and we assume that growth of these protoplanets is entirely due to the accretion of pebbles and gas from the disk. 

The requirement that the model runs quickly constrains the way certain processes can be modeled. In particular, the treatment of gas accretion and orbital interactions between multiple planets is highly simplified. For example, we assume that the orbits of the planets remain circular and coplanar, and the only dynamical interaction between them is possible capture into a mean-motion resonance.

In the following subsections, we describe the model in more detail.

%
%
\subsection{Initial Conditions}
For each optimization run, we consider a set of 2000 isolated stars with protoplanetary disks that have randomly chosen disk masses, and inner and outer disk radii. These values are chosen to match observational constraints rather than with particular theoretical models in mind. The initial disk masses are logarithmically distributed in the range 0.001 to 0.1 solar masses following \citet{williams:2011}.  (We note that this observed range of masses may not reflect the initial disk masses necessarily, but it should at least encompass the likely disk mass range at the time protoplanets are forming.) Outer disk radii $\router$ are uniformly distributed between 30 and 200 AU \citep{williams:2011}, and disk inner edges $\rinner$ are uniformly distributed between 0.02 and 0.09 AU \citep{lee:2017}. In all cases the stellar mass $\mstar$ is assumed to be one solar mass.

The initial masses of all protoplanets are the same: $M_0$, which is a model parameter. The semi-major axes of the initial protoplanets are logarithmically distributed between 0.1 and 30 AU, with the constraint that neighboring planets in multi-planet systems must be initially separated by more than their mutual 2:1 mean-motion resonance. 

All simulations are halted after 5 My.

%
%
\subsection{Gas Disk Evolution}
We model the evolution of the gas disk by assuming that the gas accretion rate onto the star at time $t$ is given by
\begin{equation}
\left(\frac{dM}{dt}\right)_{\rm gas}=\left(\frac{dM}{dt}\right)_{\rm gas,0}
\left(\frac{1}{1+t/\tdisk}\right)
\label{eq_dmdt_disk}
\end{equation}
where $\tdisk$ is the gas flux decay timescale, taken to be 0.1 million years following \citet{hartmann:2016}. The initial gas accretion rate is chosen so that the total mass of the gas disk would fall to zero after 10 million years. Given the total gas mass at time $t$, the surface density $\sigmagas$ at each distance $a$ from the star is calculated assuming that $\sigmagas$ follows a $1/a$ profile, truncated at the disk's outer edge.

We assume that the disk temperature follows a fixed profile
\begin{equation}
T=300\left(\frac{a}{\rm AU}\right)^{-1/2}\ K
\end{equation}
The maximum disk temperature is truncated at 1500 K, which is the temperature at which rocky pebbles are assumed to evaporate.

The gas is assumed to be turbulent, with a strength described by a model parameter $\alpha$. Turbulence primarily affects the dynamics of the pebbles, as described below. We assume that the gas accretion rate in the disk follows Eqn.~\ref{eq_dmdt_disk} rather than depending on the turbulence parameter $\alpha$.

%
%
\subsection{Pebble Accretion}
Pebbles drift into the planet-forming region from the outer disk, due to gas drag, at a rate given by
\begin{equation}
\left(\frac{dM}{dt}\right)_{\rm peb}=
\left(\frac{dM}{dt}\right)_{\rm peb,0}
\exp\left(-\frac{t}{\tpeb}\times\frac{\router}{\rm 100\ AU}\right)
\end{equation}
where $\tpeb$ is a model parameter. The initial pebble flux is chosen so that the total population of pebbles would be removed after infinite time. Note that we have scaled the pebble removal timescale by the radial extent of the disk since pebbles are likely to depleted in small disks more rapidly than large ones.

At each radial location in the disk, pebbles are assumed to have grown to a size at which mutual collisions become destructive. We consider two sources of relative motions between pebbles: turbulence and radial drift. We assume that pebbles have grown to the point at which the typical relative velocity equals the fragmentation velocity $\vfrag$, which a model parameter. The pebble Stokes number $\st$ is given by the smaller of the values set by turbulence \citep{ormel:2007} and radial drift \citep{weidenschilling:1977}, which are
\begin{eqnarray}
\st&=&\frac{\vfrag^2}{3\alpha c_s^2} 
\hspace{20mm} {\rm turbulence} \nonumber \\
&=&\frac{\vfrag}{\eta\vkep}
\hspace{20mm} {\rm drift}
\end{eqnarray}
where $\vkep$ is the Keplerian velocity, $c_s$ is the gas sound speed, $\eta\simeq(c_s/\vkep)^2$ is the fractional difference between the orbital speed of the gas and a solid body, and $\alpha$ is a model parameter that provides a measure of the turbulence strength. Note that in deriving the relative velocity of pebbles due to radial drift, we have assumed that the typical relative velocity for pebbles with a limited size spectrum is half the absolute drift velocity.

Given the pebble Stokes number and mass flux, the pebble surface density in the planet forming region is given by
\begin{equation}
\sigmapeb=\frac{1}{2\pi av_r}\times\left(\frac{dM}{dt}\right)_{\rm peb}
\end{equation}
where the radial drift velocity $v_r$ is given by
\begin{equation}
v_r=\frac{2\eta\vkep\st}{1+\st^2}
\end{equation}
\citep{weidenschilling:1977}.

The initial pebble-to-gas mass ratio in the disk is 0.01 in all cases. Pebble surface densities drop by a factor of 2 inside the ice line at 160 K. For a disk with outer radius of 100 AU, and typical model parameters found by the optimization scheme, the initial pebble-to-gas surface density ratio is around 0.01, falling by roughly an order of magnitude after 1 My. The initial pebble-to-gas mass flux ratio for the same parameters is about 0.02, falling to about 0.007 after 1 My. Essentially all the pebbles in the disk are removed by drift or accretion by the end of a simulation.

We calculate the rate at which a planet accretes pebbles following \citet{ormel:2010}. The mass $\mplan$ of the planet grows at a rate
\begin{eqnarray}
\left(\frac{d\mplan}{dt}\right)_{\rm peb}
&=&2\rcap\vrel\sigmapeb
\hspace{20mm} \rcap>\hpeb \nonumber \\
&=&\frac{\pi\rcap^2\vrel\sigmapeb}{2\hpeb}
\hspace{20mm} \rcap<\hpeb
\end{eqnarray}
where $\rcap$ is the capture radius for pebbles, given by
\begin{equation}
\left(\frac{\rcap}{r_H}\right)^3
+\frac{2a\vrel}{3r_H\vkep}\left(\frac{\rcap}{r_H}\right)^2-8\st=0
\end{equation}
and $r_H$ is the planet's Hill radius, given by
\begin{equation}
r_H=a\left(\frac{\mplan}{3\mstar}\right)^{1/3}
\end{equation}

In addition, $\hpeb$ is the scale height of the pebbles due to turbulence \citep{youdin:2007}, given by
\begin{equation}
\hpeb=\hgas\left(\frac{\alpha}{\alpha+\st}\right)^{1/2}
\end{equation}
and $\vrel$ is the relative velocity between the planet and a pebble, given by
\begin{equation}
\vrel=\vkep\times\max\left[\eta,\frac{3\rcap}{2a}\right]
\end{equation}

Note that when more than one planet is present, the local flux of pebbles seen by a given planet is reduced as a result of pebbles accreted by more distant planets. In extreme cases, this means that the supply of pebbles to inner planets can be shut off entirely. 

Following \citet{lambrechts:2014}, we also assume that massive planets can stop the inward flow of pebbles due to their gravitational effect on the nearby radial profile of the gas pressure. We assume that a planet is large enough to terminate the flux of pebbles to itself and any planets closer to the star when its mass exceeds $\miso$ given by 
\begin{equation}
\miso=\fiso\mgap
\end{equation}
where $\mgap$ is the planetary mass needed to open a gap in the gas disk, described below, and $\fiso$ is a model parameter.

%
%
\subsection{Gas Accretion}
Low-mass planets embedded in a protoplanetary disk capture an atmosphere from the disk.  Here, we assume that these atmospheres are nearly static, with a balance between the inward pull of the planet's gravity and the outward pressure gradient within the atmosphere. The atmosphere is heated and prevented from collapsing by the energy supplied by infalling solid particles, here assumed to be pebbles. For planets above a certain mass, the luminosity generated by infalling solids is insufficient to balance gravity, and gas begins to flow inwards \citep{mizuno:1980, stevenson:1982}. The atmosphere contracts slowly on a cooling timescale, with the luminosity supplied by infalling gas restoring an approximate balance with gravity \citep{piso:2014}. (We note that some recent hydrodynamic simulations of super-Earth-mass planets in protoplanetary disks suggest that advection of disk gas through the outer atmosphere can substantially affect gas accretion rates \citep{ormel:2015, lambrechts:2017}. We do not consider this effect here.)

For a planet of mass $\mplan$ that is massive enough to accrete gas, we use the following expression for the maximum gas accretion rate given an unlimited supply of gas:
\begin{equation}
\left(\frac{d\mplan}{dt}\right)_{\rm gas,\ max}
=
\left(\frac{\mplan}{5M_\oplus}\right)^{7/2}
\left(\frac{\mcore}{2.5M_\oplus}\right)^{-1}
\left(\frac{\matmos}{2.5M_\oplus}\right)^{-1}
\left(\frac{T}{160\ K}\right)^{-1/2}
\frac{\mplan}{\tcool}
\end{equation}
where $T$ is the local disk temperature, $\mcore$ is the mass of the solid portion of the planet, $\matmos$ is the mass of the planet's atmosphere, and $\tcool$ is a model parameter. (For very low mass planets we set a minimum value for $\matmos$ of $0.01M_\oplus$ in this expression.) The form of this equation is justified in Appendix~A, and is broadly similar to the expression for gas accretion used by \citet{bitsch:2015}.

We assume that mass in the atmosphere is concentrated near the planet's solid core \citep{lee:2015}. As a result, the luminosity generated by a unit mass of infalling gas is approximately the same as that generated by a unit mass of infalling pebbles, assuming pebbles also fall to the core. Thus, the actual gas accretion rate is given by
\begin{equation}
\left(\frac{d\mplan}{dt}\right)_{\rm gas}=
\max\left[0,
\left(\frac{d\mplan}{dt}\right)_{\rm gas,\ max}
-\left(\frac{d\mplan}{dt}\right)_{\rm peb}
\right]
\end{equation}

The maximum possible gas accretion rate is set by the mass flux of gas flowing inwards through the disk minus any gas flux accreted by more distant planets. In practice, we assume that a planet can only accrete a fraction $\fgas$ of the disk gas flux, where $\fgas$ is a model parameter.

%
%
\subsection{Orbital Migration}
Planetary orbits can migrate due to tidal interactions with the disk. For low-mass planets, we calculate migration using the formula developed by \citet{tanaka:2002} for linear, type-I migration in isothermal disks. The distance $a$ from the star changes at a rate
\begin{equation}
\left(\frac{da}{dt}\right)_{\rm type\ I}=-3.8
\left(\frac{\mplan}{\mstar}\right)
\left(\frac{\sigmagas a^2}{\mstar}\right)
\left(\frac{\vkep}{c_s}\right)^2
\vkep
\label{eq_type1}
\end{equation}
where the constant of proportionality is appropriate for a disk with surface density $\sigmagas\propto1/a$.

Planets above a certain mass begin to open a gap in the gas disk. This affects the migration rate, and also the inward drift of pebbles discussed above. We use the gap opening formulae derived by \citet{rafikov:2002}, where the minimum planetary mass needed to open a gap is given by
\begin{equation}
\mgap=\max\left[\minv,\mvis\right]
\end{equation}
where
\begin{eqnarray}
\minv&=&M_1\times\min\left[5.2Q^{-5/7}, 
3.8\left(\frac{Qa}{\hgas}\right)^{-5/13}\right]
\nonumber \\
\mvis&=&M_1\times\left(\frac{a\alpha}{0.043\hgas}\right)^{1/2}
\end{eqnarray}
where $\hgas$ is the gas scale height, and
\begin{equation}
Q=\frac{c_s\vkep}{\pi aG\sigmagas}
\end{equation}
and
\begin{equation}
M_1=\frac{2ac_s^3}{3G\vkep}
\end{equation}

Planets close to or above the gap-opening mass are likely to migrate more slowly than Eqn.~\ref{eq_type1} would suggest since much of the disk material close to the planet has been removed, weakening the tidal interactions. Early studies predicted that a planet's migration would become locked to the overall flow of gas at this point, so-called type-II migration \citep{ward:1997}. However, recent simulations failed to reproduce this effect, and found instead that massive planets migrate at a rate determined solely by the applied torques \citep{duffell:2014, durmann:2015}. Here, we assume that type-I migration also applies to massive planets, but at a reduced rate due to the weakening of torques as gas is cleared from the gap. The more massive the planet, the deeper the gap will be, and the greater the reduction in the migration rate. For this reason, we replace $\mplan$ in Eqn.~\ref{eq_type1} with an effective mass $\meff$ given by
\begin{eqnarray}
\meff&=&\mplan \hspace{45mm} \mplan<0.5\mgap
\nonumber \\
&=&0.5\mgap\times\left(\frac{\mplan}{0.5\mgap}\right)^x
\hspace{20mm} \mplan>0.5\mgap
\label{eq_type1_mod}
\end{eqnarray}
where $x$ is a model parameter. For example, $x=1$ corresponds to standard type-I migration for all planetary masses; while for $x<0$, the migration speed decreases with increasing mass. Migration rates relative to Eqn.~\ref{eq_type1} are likely to decline with increasing mass as the gap cleared by the planet grows deeper, but the precise relationship remains uncertain. The form of Eqn.~\ref{eq_type1_mod} reflects the expected overall trend while adding only a single new model parameter to accommodate the uncertainty.

Many of the cases described in this paper consider multi-planet systems. When planets migrate, a pair of planets with converging orbital periods can be captured into a mean-motion resonance. Here, we assume that neighboring planets will always be captured in the 2:1 mean-motion resonance if their period ratio falls below 2, independent of planetary mass. These planets remain in the resonance as long as their nominal migration rates do not cause their periods to diverge. Additional planets can be captured into resonance with one of these bodies, forming resonance chains of 3 or more planets. When treating these objects, we calculate the total loss of angular momentum for the chain due to migration for each object, and then calculate new orbits assuming that neighboring planets maintain a 2:1 period ratio.

More complicated models for orbital interactions and resonance capture could certainly be applied here, but we adopt this simple scheme in order to keep the model as efficient as possible.

%
%
\subsection{Model Parameters}
The model described above contains 8 adjustable parameters that have large uncertainties. These parameters are listed in Table~1 together with the range of allowed values that we consider here.

\startlongtable
\begin{deluxetable}{lll}
\tablecaption{Model Parameters to be Optimized}
\tablehead{
\colhead{Parameter} & \colhead{Symbol} & \colhead{Range} \\
}
\startdata
Disk gas turbulence parameter & $\alpha$ & $10^{-5}$--$10^{-2}$ \\
Pebble flux decay timescale for 100 AU radius disk & $\tpeb$ & $3\times10^5$--$3\times10^7$ y \\
Initial planet mass & $M_0$ & $10^{-7}$--$3\times 10^{-3} M_\oplus$ \\
Pebble isolation mass / disk gap-opening mass & $\fiso$ & 0.01--1 \\
Pebble fragmentation speed & $\vfrag$ & 1--1000 cm/s \\
Gas cooling timescale for 5 $M_\oplus$ planet & $\tcool$ & $10^6$--$10^9$ y \\
Maximum gas accretion rate / disk gas flux & $\fgas$ & 0.01--1 \\
Modified migration rate planetary mass exponent & $x$ & -2 to 1 \\
\enddata
\end{deluxetable}

%
%
\section{Observational Constraints} \label{sec:obs}
\begin{figure}
\plotone{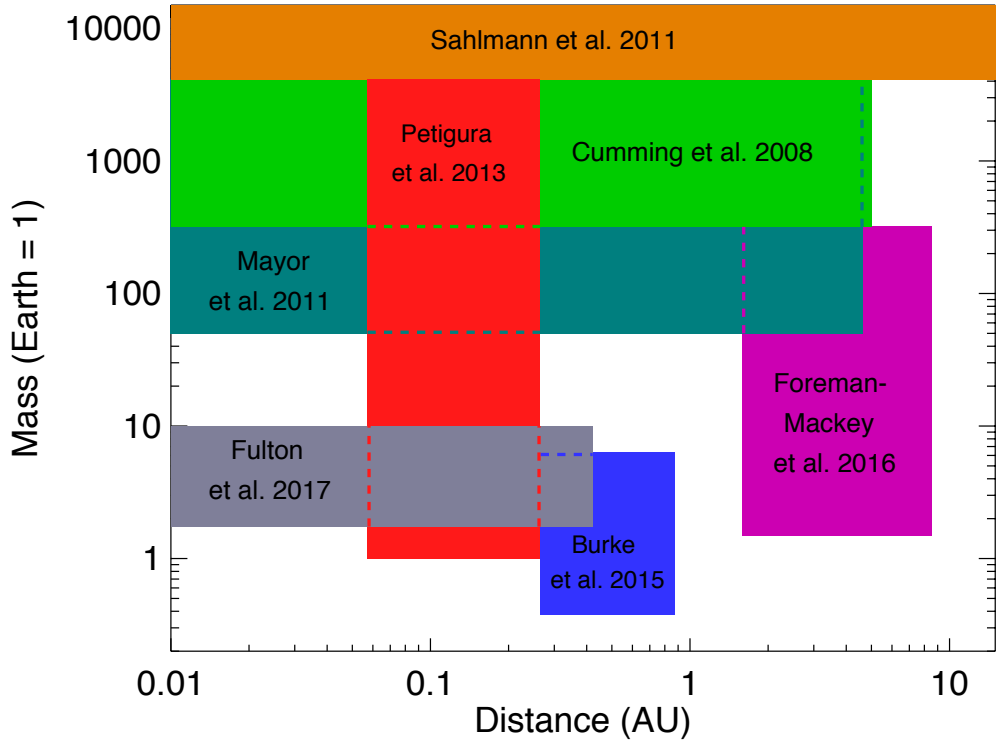}
\caption{Extent of the regions covered by each of the observational surveys in planetary mass-orbital distance space. For the purpose of this figure only, we assume that planetary radii and masses are related by the empirical relation found by \citet{weiss:2014}.}
\end{figure}

In this section we describe the observational constraints that are used to quantify the success of the simulations described above. We used published observational estimates for the frequency of planets with a variety of orbital periods and masses/radii, as well as the frequency of brown dwarfs. The constraints were obtained from surveys for extrasolar planets using the Doppler radial velocity and transit photometry methods. The constraints are listed in the following subsections.

Figure~1 shows the extent of each observational survey in planetary mass-orbital distance phase space. For the purpose of this figure only, we assume that planetary masses and radii are related using the empirical relation found by \citet{weiss:2014}. In the actual runs described below, we calculate planetary radii based on planetary composition, and compare with observed radii or masses as appropriate. The observational constraints cover most of the phase space interior to 10 AU and above 1 Earth mass.

\subsection{\citet{petigura:2013}} This study used data from the Kepler mission to estimate the fraction of Sun-like stars that have a planet with a radius at least as large as Earth. The planetary frequencies obtained by the authors are divided into bins according to orbital period $P$ as follows
\begin{itemize}
\item $5.0<P<7.3$ days: $1.6\pm0.4$\% of stars have a planet.
\item $7.3<P<10.8$ days: $2.7\pm0.7$\% of stars.
\item $10.8<P<15.8$ days: $4.6\pm1.0$\% of stars.
\item $15.8<P<23.2$ days: $5.2\pm1.3$\% of stars.
\item $23.2<P<34.1$ days: $5.0\pm1.3$\% of stars.
\item $34.1<P<50.0$ days: $5.8\pm1.7$\% of stars.
\end{itemize}

\subsection{\citet{mayor:2011} and \citet{santerne:2016}} The study by \citet{mayor:2011} estimated the fraction of F, G and K stars with planets with masses $M>50M_\oplus$ and periods less than 10 years, found using the HARPS and CORALIE radial velocity search. In a subsequent work, \citet{santerne:2016} analyzed the same data and estimated the fraction of stars with similarly massive planets for several other period ranges. The estimated planetary frequencies from these studies are
\begin{itemize}
\item $P<10$ days: $0.83\pm0.34$\% of stars have a planet.
\item $10<P<85$ days: $1.64\pm0.55$\% of stars.
\item $85<P<400$ days: $2.90\pm0.72$\% of stars.
\item $P<10$ years: $13.9\pm1.7$\% of stars,
\end{itemize}

\subsection{\citet{cumming:2008}} This study estimated the fraction of F, G and K stars with planets greater than or equal to Jupiter in mass, using 8 years of radial velocity data from the Keck Planet Search. The planetary frequencies were binned by orbital period as follows
\begin{itemize}
\item $P<11.5$ days: $0.43\pm0.30$\% of stars have a planet.
\item $P<100$ days: $0.85\pm0.40$\% of stars.
\item $P<365$ days: $1.9\pm0.6$\% of stars.
\item $P<1022$ days: $3.9\pm0.9$\% of stars.
\item $P<1896$ days: $4.6\pm1.0$\% of stars.
\item $P<4080$ days: $8.9\pm1.4$\% of stars.
\end{itemize}

\subsection{\citet{fulton:2017}} This work estimated the number of planets per star with periods less than 100 days using data from the California Kepler Survey for stars with temperatures in the range 4700--6500 K. The planetary frequencies were binned by planetary radius as follows
\begin{itemize}
\item $1.15<R/R_\oplus<1.29$: $0.078\pm0.017$ planets per star.
\item $1.29<R/R_\oplus<1.43$: $0.080\pm0.013$ planets.
\item $1.43<R/R_\oplus<1.59$: $0.053\pm0.011$ planets.
\item $1.59<R/R_\oplus<1.77$: $0.0334\pm0.0092$ planets.
\item $1.77<R/R_\oplus<1.97$: $0.050\pm0.010$ planets.
\item $1.97<R/R_\oplus<2.19$: $0.086\pm0.016$ planets.
\item $2.19<R/R_\oplus<2.43$: $0.098\pm0.016$ planets.
\item $2.43<R/R_\oplus<2.70$: $0.077\pm0.016$ planets.
\item $2.70<R/R_\oplus<3.00$: $0.053\pm0.012$ planets.
\item $3.00<R/R_\oplus<3.33$: $0.0316\pm0.0089$ planets.
\item $3.33<R/R_\oplus<3.70$: $0.0242\pm0.0066$ planets.
\item $3.70<R/R_\oplus<4.12$: $0.0094\pm0.0057$ planets.
\end{itemize}
Note that this study calculated the number of planets per star, which is not necessarily the same as the fraction of stars with such a planet because some stars may have more than one planet.

\subsection{\citet{burke:2015}} This study estimated the number of planets per star with period $50<P<300$ days, and radius between 0.75 and 2.5 Earth radii, for G and K stars observed by the Kepler mission. The authors estimated that there are $0.77$ such planets per star, with an allowed range of 0.28 to 1.9. This range includes statistical and systematic uncertainties. In general, our model tends to produce values of this statistic lower than 0.77. For the purposes of calculating how well the model fits matches observations we will assume that the allowed range corresponds to a $2\sigma$ uncertainty, given by $0.77-0.28$. Thus we assume that there should be $0.77\pm0.245$ planets per star within this period and radius range.

\subsection{\citet{foreman:2016}} This study estimated the number of long-period planets per star for G and K stars, using data from the Kepler mission for stars exhibiting only one or two transits. The authors estimate that there are $2.00\pm0.72$ planets per star with periods between 2 and 25 years, and radii between 0.1 and 1 Jupiter radius.

\subsection{\citet{sahlmann:2011}} This study estimated the fraction of stars that have a brown dwarf  companion with a mass between 13 and 80 Jupiter masses, using data obtained by the CORALIE radial velocity survey. The authors show that out of about 1600 stars surveyed, 11 have plausible brown dwarf companions, equivalent to  $0.69\pm0.21$\% of stars.

Note, in this study, we only consider the formation of brown dwarfs by core accretion rather than other formation pathways. However, as we will see, the model is able to match the observed brown dwarf population easily in a variety of cases, so the fraction of brown dwarfs that form via core accretion may not provide a strong constraint on the kind of planet formation model used here.

%
%
\section{Optimizing the Model Parameters} \label{sec:opt}
The model described in Section~\ref{sec:model} has 8 free parameters that are listed in Table~1. In order to find the best values for these parameters, we ran the model many times, and compared the distribution of the resulting planets with the observationally derived distributions listed in Section~\ref{sec:obs}. Based on the fit, we obtained improved parameter values that were used to perform new runs, and so on. This iterative process was automated using a particle swarm optimization scheme \citep{poli:2007} in order to find the best solution in the 8-dimensional parameter space.

Prior to the start of an optimization run, we generated $\nsys$ sets of initial conditions, where $\nsys=2000$. Each set of initial conditions consists of the protoplanetary disk mass, inner and outer disk radii, and initial orbital distance(s) for the protoplanet(s).

During the optimization procedure we considered a small number $\npart$ (typically 40) of possible solutions called ``particles''. Each particle is an 8-dimensional vector $\bf{x}$ where the components of the vector consist of values for each of the 8 model parameters. At each optimization step, and for each particle, we simulate the formation of $\nsys$ planetary systems, using the same set of model parameters but different initial conditions for each system. Thus, for each optimization step we simulate the formation of $\npart\times\nsys$ planetary systems.

After running the simulations, we calculate a score for each particle by comparing the distribution of simulated planets with the observational constraints described in Section~\ref{sec:obs}. The precise way in which the score is calculated is described in more detail below. For each particle $i$, the particle vector $\bf{x}_i$ is then updated using
\begin{eqnarray}
{\bf v}_i&=&C_1{\bf  v}_i+C_2z_1({\bf p}_i-{\bf x}_i)+C_3z_2({\bf g}_i-{\bf x}_i)
\nonumber \\
{\bf x}_i&=&{\bf x}_i+{\bf v}_i
\label{eq-pso}
\end{eqnarray}
where $\bf{v}_i$ can be thought of as a velocity in the 8-dimensional phase space. Here, $z_1$ and $z_2$ are uniform random numbers between 0 and 1. The vector ${\bf p}_i$ represents the set of model parameters for the best solution found so far {\em by this particle} (the ``personal best''). The vector ${\bf g}_i$ represents the best solution found so far {\em by any particle} (the ``global best''). We set the optimization parameters to $C_1=0.83$ and $C_2=C_3=1.65$. The initial values for the ${\bf x}_i$ are chosen at random within the ranges listed in Table~1. During optimization, we allow the particles to move outside this range, but scores are only calculated for particles when they lie within this 8-dimensional box.

Using this particle swarm optimization (PSO) scheme, each particle randomly explores phase space, but tends to follow a damped orbit about the best solution it has found so far. At the same time, the particle is accelerated towards the best solution found by any particle, so that the particle occasionally discovers and starts orbiting a new, better solution of its own. The inertia parameter $C_1$ ensures that particles do not stray too far afield and eventually converge towards a single solution. The stochastic nature of PSO helps a system find the global minimum of a function (in this case the score) in a phase space that contains multiple local minima \citep{poli:2007}.

We ran the optimization scheme for $\niter$ iterations, where $\niter=5000$. At each iteration, we use the same set of $\nsys$ initial conditions, since the best fit solution can be somewhat  different for different sets of initial conditions. When finished, an optimization run will have simulated the formation of $\niter\times\npart\times\nsys\sim10^8$ planetary systems, which explains the need for a relatively simple and efficient planet-formation model.

Readers interested in more details of the PSO runs and tests for the convergence of the PSO scheme can consult Appendix~B.

The score used by the optimization scheme is calculated as  follows. We first note the orbital distances $a$ and masses $M$ of the planets in each system at 5 My. The orbital periods are determined using Kepler's third law. The planetary radii $R$ are calculated from the total mass and the gas fraction using the empirical formulae for evolved planets developed by \citet{lopez:2014}. The score $Z$ for each particle is calculated using
\begin{equation}
Z^2=\sum_{j=1}^{31}W_j\left(\frac{y_j-\mu_j}{\sigma_j}\right)^2
\end{equation}
where $y_j$ and $\mu_j$ are the simulated and observed values for each of the 31 constraints listed in Section~\ref{sec:obs}, and $\sigma_j$ is the published error for each constraint. The weights $W_j$ are chosen so that each of the 7 sets of constraints (the 7 observational studies listed in Section~\ref{sec:obs}) are weighted equally, with the 31 weights summing to unity.

%
%
\section{The 1-Planet Case} \label{sec:n1}
\begin{figure}
\plotone{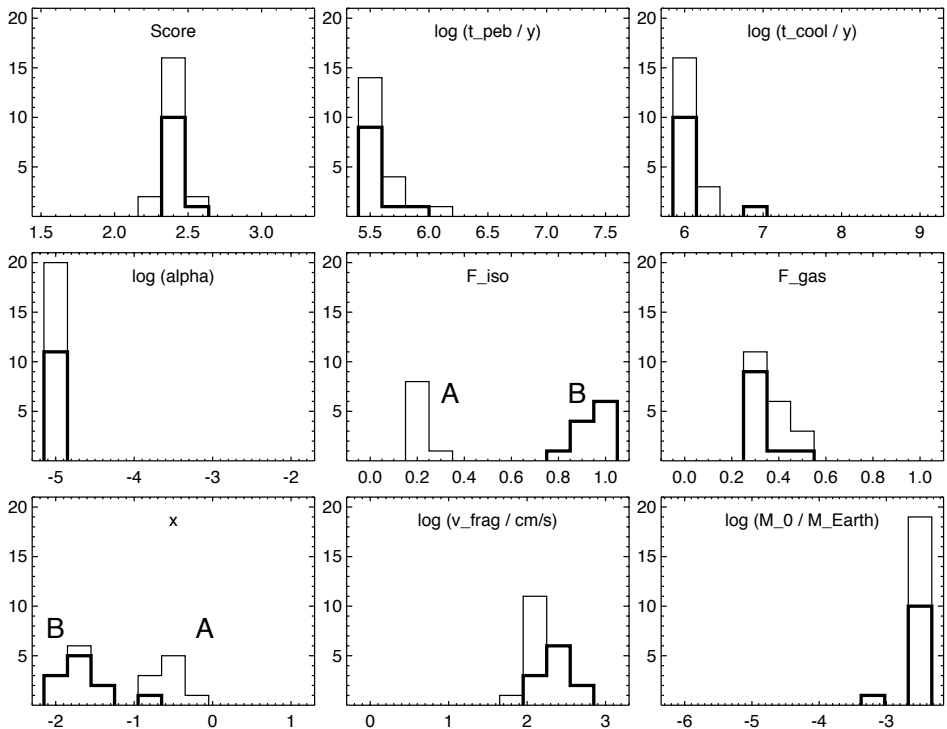}
\caption{Distribution of scores and best-fit model parameters for 20 optimization runs for systems with a single planet per star. The best fit solutions are clustered into two groups labelled A and B.}
\end{figure}

We begin by looking at systems that contain a single planet per star. This case is commonly used in population synthesis studies \citep{ida:2004, ida:2008, mordasini:2009b}, and has the advantage that no model is required for planetary interactions.

We ran the optimization scheme 20 times using different random number seeds for the model initial conditions and PSO particles. Figure~2 shows the final scores for the 20 cases, and the distribution of best-fit model parameters. Nine of the 20 cases have converged to essentially the same solution, with some scatter due to the different initial conditions in each case. We refer to this as Solution~A below. The remaining 11 cases (marked by the histogram with the bold outline) found a second solution (Solution~B) with a slightly higher score (poorer fit) and different parameter values.

The scores for all 20 runs are quite large---typically about 2.4 standard deviations away from the mean  observational  constraints on average.  We also ran cases with different fragmentation speeds (and thus different sizes) for pebbles interior and exterior to the ice line, inspired by the work of \citet{morbidelli:2015}, without achieving a significant improvement in the score. The tight clustering of the scores in Figure~2 suggests this is about as good as the simple planet-formation model can achieve for the 1-planet case, at least for the range of parameter values considered here. In the next section, we will see that the fit can be improved by adding more planets.

We can draw some conclusions from the best-fit values for the 8 model parameters. The values for the pebble drift timescale $\tpeb$ cluster near the smallest permitted value of $3\times10^5$ years for a 100 AU radius disk. For most of the first My, the gas and pebble fluxes decline at similar rates in a 100 AU disk, leaving the dust-to-gas ratio very roughly constant. At later times the pebble flux declines more rapidly, giving progressively lower dust-to-gas ratios. For this value of $\tpeb$, the pebbles have essentially disappeared by 2--3 My for all disk radii considered here. Thus, the growth of the solid portion of planets mostly occurs at early times.

\begin{figure}
\plotone{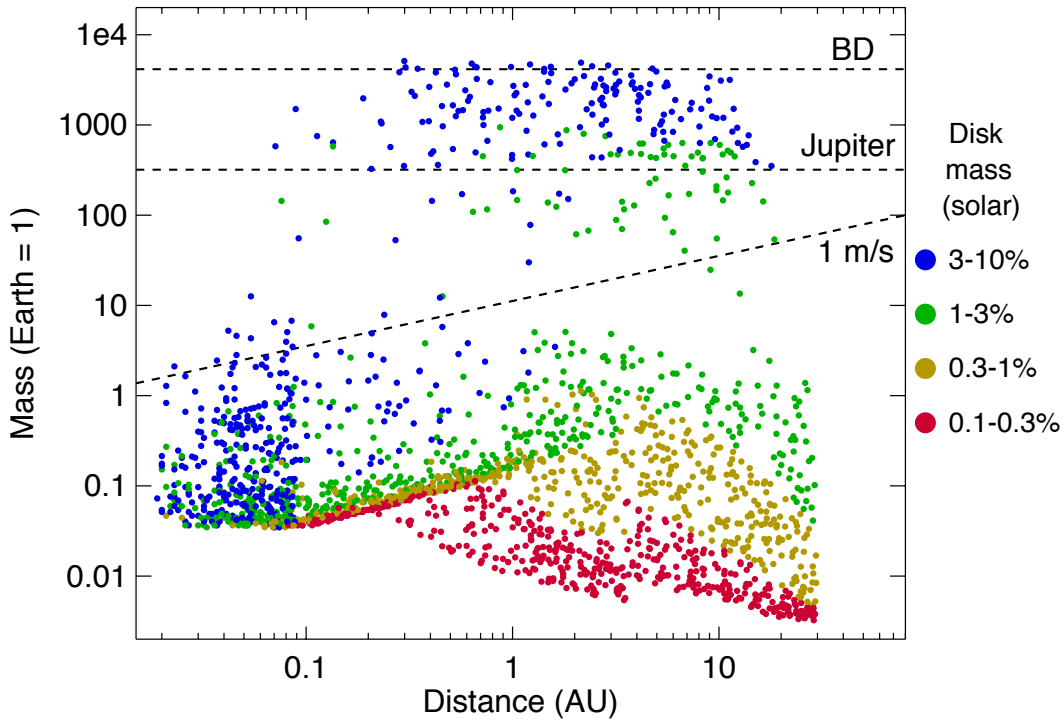}
\caption{Final masses and orbital distances of planets in 2000 systems, each with a single planet, for the best fit model. The upper two dashed lines indicate 1 and 13 Jupiter masses (the latter labelled ``BD'' for brown dwarf). The lowest dashed line indicates a planetary mass that generates a 1 m/s Doppler radial velocity signal in the star. Different symbol colors indicate planets that formed in disks with different initial masses.}
\end{figure}

The disk turbulence levels $\alpha$ are typically close to $10^{-5}$, the lowest value permitted. In the model used here, the main role of turbulence is to set the scale height for pebbles in the disk. Low values for $\alpha$ lead to a thin pebble layer, which leads to rapid growth of low-mass planets due to pebble accretion. The best-fit initial protoplanet mass $M_0$ tends to be near the upper permitted value, and this also favors rapid growth, since pebble accretion rates increase rapidly with planetary mass for low-mass objects \citep{lambrechts:2012}.

The best-fit solutions find pebble fragmentation speeds $\vfrag$ of about 1--3 m/s. These values are similar to experimental results that find that collisions between rocky particles tend to become erosional at speeds above about 1 m/s \citep{guttler:2010}. Some work has suggested that pebble growth may stall at much smaller sizes if collisions lead to bouncing rather than sticking \citep{zsom:2010}. This possibility is not supported by the relatively large values of $\vfrag$ found here.

The point at which planets become large enough to halt the inward drift of pebbles is controlled by the parameter $\fiso$. The two solutions found by the optimization scheme show a large difference in $\fiso$. Solution~B favors $\fiso\simeq1$, which means that planets continue to accrete pebbles until they are large enough to clear a deep gap in the gas disk. Solution~A finds that pebble accretion ceases at much lower masses, when the planet is only massive enough to modestly affect the radial profile of the gas in its vicinity. Hydrodynamical simulations of pebble drift and accretion tend to find isolation masses that are lower than the gap opening mass, qualitatively supporting the latter solution, although these studies typically consider disks with much higher turbulent viscosities \citep{lambrechts:2014}. Recently, \citet{bitsch:2018b} have found a trend of decreasing pebble isolation mass with decreasing disk viscosity, although the isolation masses are larger than those predicted here.

The atmospheric cooling times $\tcool$ cluster near the lowest permitted value. For planets near the critical mass for the onset of gas accretion, the rate at which the atmosphere cools is mainly determined by the dust opacity \citep{lee:2015}. The short cooling times found here favor low dust opacities, in agreement with recent studies that suggest dust tends to form large aggregates in planetary atmospheres, reducing the opacity \citep{movshovitz:2008, mordasini:2014}.

The favored migration rates for high-mass planets are also somewhat different for the two solutions. In the model used here, planets that are massive enough to open a gap in the gas disk migrate more slowly than the type-I migration exhibited by low-mass planets. The modified migration rate is proportional to $\mplan^x$ where $\mplan$ is the total mass of the planet and $x$ is an optimized parameter. Both best-fit solutions find $x<0$, so that migration rates decrease with increasing mass. However, the reduction in migration rate with mass is greater for Solution~B than for Solution~A. The values of $x$ are broadly distributed, but center around $x=-1$. This is the same dependence often used in models for type-II migration in cases where the planetary mass is comparable to or greater than that of the disk \citep{mordasini:2009a, hasegawa:2013}.

Finally, the maximum gas-accretion efficiencies $\fgas$ cluster around 0.4. Thus, very massive planets are able to accrete only 40\% of the gas in their vicinity, while the remainder of the gas flows past the planet and crosses the gap in the disk generated by the planet.

\begin{figure}
\plotone{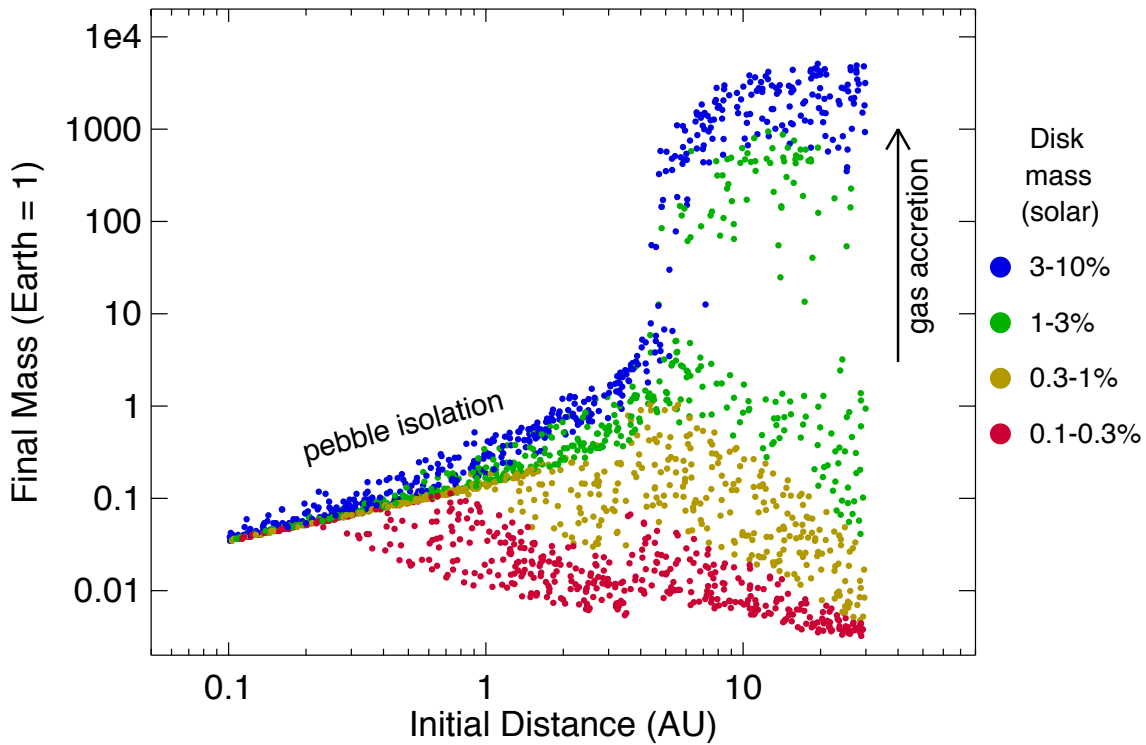}
\caption{Final planetary masses as a function of {\em initial\/} orbital distance for 2000 single-planet systems. Different symbol colors indicate planets that formed in disks with different initial masses. Planets initially inside 4 AU never grow above the pebble isolation mass, which is too low for the planet to undergo significant gas accretion. Planets that start beyond 4 AU in massive disks can become large enough to accrete gas and form giant planets.}
\end{figure}

Table 2 shows the frequency of planets produced by two of the best-fit models compared  to the observations for the 31 regions of mass-period (or radius-period) space described in Section~3. One case is an example of Solution~A, the other an example of Solution~B. The table also shows the discrepancy between model and observations in terms of the estimated error $\sigma$ on the observations (columns labelled $N\sigma$). 

The single-planet model clearly does better at matching some of the observations than others. The fit for both solutions is substantially better for giant planets and brown dwarfs than for low-mass planets. The model consistently produces too few small planets compared to observations, often by a large amount. The observed frequency of brown dwarfs is matched almost perfectly for both solutions, which suggests this is straightforward to achieve. As a result, reproducing the brown dwarf frequency may not be a strong test of models for planet formation.

The reason for the difference between the model and observations can be seen more clearly in Figure~3, which shows the final orbits and masses for the planets for a Solution~A case. The majority of protoplanets failed to grow larger than 1 Earth mass, which means they would not be counted by the observational surveys described in Section~3. Super-Earths, with masses between 1 and 10 Earth masses are not particularly abundant in the model, whereas they are very common in observational surveys. 

Almost all of the model planets that grew larger than Earth formed in disks with above average mass. This can be seen by comparing the red and yellow points in Figure~3, which indicate planets formed in disks with initial mass $\mdisk<0.01M_\odot$, and the blue and green dots, which are planets formed in disks with $\mdisk>0.01M_\odot$. Note that there are roughly equal numbers of objects in each category. It makes sense that planets should grow larger in more massive disks, since more solid material is available, but this is not the only factor at work, as we will see below.

One observed feature that is reproduced by the model is a gap in the mass distribution between about 10 and 50 Earth masses. This results from the rapid rate of gas accretion for objects in this mass range. Qualitatively similar features have been seen in some other population synthesis studies \citep{ida:2004, ida:2008}. Another clear feature of Figure~3 is the pile up of low mass planets between 0.02 and 0.09 AU. These are objects that have migrated inwards until they reach the inner edge of the disk, at which point migration ceases.
 
Another way to examine the planets formed in the model is to view the final planetary masses as a function of the planets {\em starting location}, as shown in Figure~4. Plotting the data in this way disentangles the effects of planetary growth and migration. In Figure~4, the planets interior to about 4 AU show a clear upper limit to growth which increases slowly with distance from the star. This upper limit is the pebble isolation mass---the mass of a planet at which it perturbs nearby gas sufficiently to halt the supply of pebbles drifting inwards. Essentially all planets inside 4 AU in massive disks grow fast enough to reach the pebble isolation mass within the disk lifetime. Growth of the solid planet ceases at this point. Further growth can only take place by accreting gas, but planets that start inside about 4 AU are too small to accrete significant amounts of gas, so they have  reached  their final mass at this point. 

\begin{figure}
\plotone{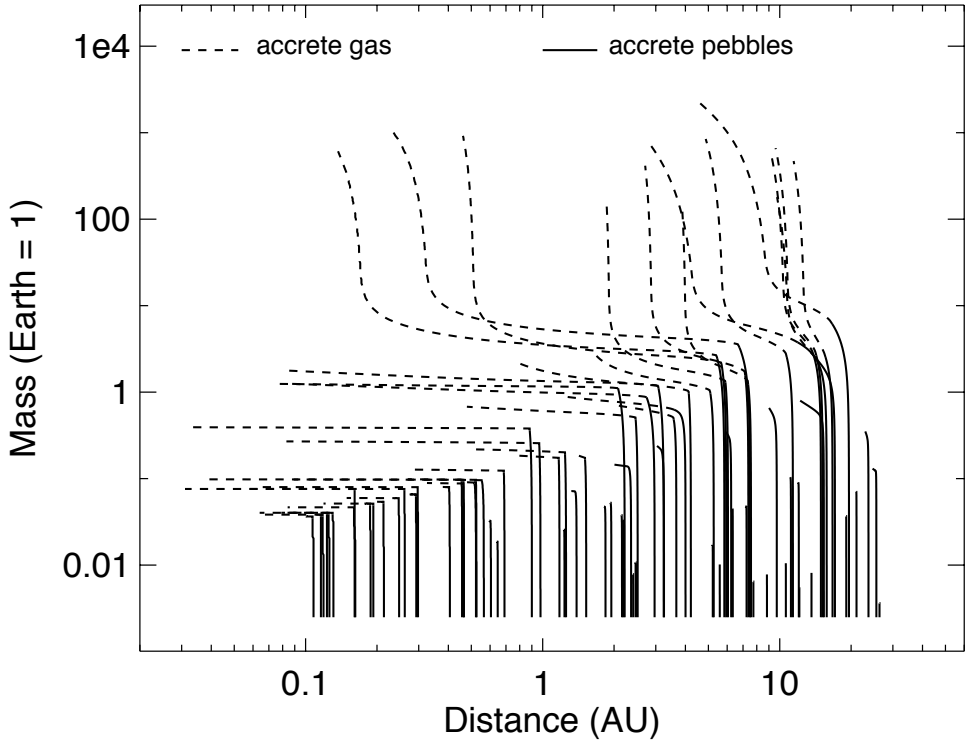}
\caption{Evolutionary tracks in mass-orbital distance space for planets in 80 single-planet systems, for the best fit model. Solid curves show epochs when planets are primarily accreting mass in the form of pebbles. Dashed curves show epochs when planets are mainly accreting gas.}
\end{figure}

Beyond about 4 AU, the pebble isolation mass is high enough that solid planets that reach this mass can begin gas accretion and become giant planets. However, only a fraction of planets beyond 4 AU grow fast enough to reach the pebble isolation mass. Many objects beyond 4 AU fail to become giants because they grow too slowly. Conversely, objects inside 4 AU fail to become giants because the pebble isolation mass is too small. Planets beyond 4 AU that grow enough to accrete gas undergo differing degrees of migration depending on their mass and the gas surface density at this stage. As a result, the final orbits of these giants populate a wide range of distances in Figure~3, extending to the inner edge of the disks.

In the model used here, the pebble isolation mass is given by $\fiso\times\mgap$ where $\fiso$ is one of the parameters fitted by the optimization scheme, and $\mgap$ is the mass needed to open a gap in the gas disk. In the inner regions of low-mass disks, $\mgap$ is set by viscosity (parameterized by $\alpha$), and is independent of gas surface density $\sigmagas$. This explains why many of the planets inside 1 AU in low-mass disks (red and yellow points) lie along a single line in Figure~4. For larger distances or larger disk masses, $\mgap$ depends on $\sigmagas$. As $\sigmagas$ increases, a planet migrates faster, so it must be more massive to clear a gap quickly before it migrates out of that region \citep{rafikov:2002}. Thus, the pebble isolation mass increases with disk mass, explaining the spread of masses for planets inside 4 AU in massive disks (blue points in Figure~4).

For the disk model used here, pebble accretion becomes less efficient with increasing distance from the star.  Below the pebble isolation mass planets grow more slowly further from the star, even though the pebble flux is independent of distance. This explains the trend with distance seen in the lower boundary of the final planetary masses in Figure~4. The small jump in this trend at about 3.5 AU occurs at the disk ice line, where the icy component of inward drifting pebbles evaporates. 

For very low mass disks (red points), planets grow very little in the outer disk, and the final masses are only slightly higher than the initial mass. In most cases, planets in low-mass disks are too small, and the gas surface densities too low, for substantial migration to occur. Thus, the distribution of red and yellow points in Figure~4 is similar to Figure~3. Planets in massive disks can migrate large distances, and the blue points in the two figures have a very different distribution.

Figure 5 shows evolutionary tracks in mass-distance space for 80 of the planets in Figure~3. The solid portions of each curve show the mass and orbital distance of a planet that is below the pebble isolation mass, and that is growing by pebble accretion. The dashed portion of each curve shows the evolution for an object above the pebble isolation mass. These objects can accrete gas, although the growth rates are very small for bodies less than an Earth mass.

\begin{figure}
\plotone{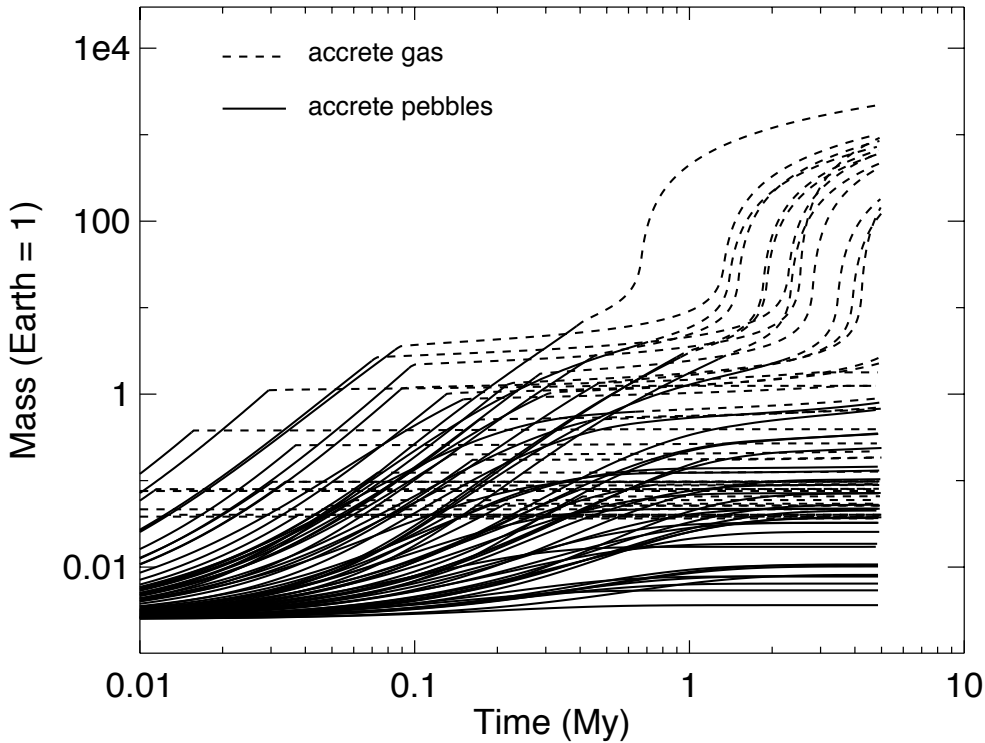}
\caption{Mass versus time for the planets in 80 single-planet systems shown in Figure~5. Solid curves show epochs when planets are primarily accreting mass in the form of pebbles. Dashed curves show epochs when planets are mainly accreting gas.}
\end{figure}

The solid curves in Figure~5 are almost vertical. Typically, planets below the pebble isolation mass are too small to migrate rapidly for the value of $\fiso$ here, The only exceptions are planets beyond about 10 AU. Here the pebble isolation mass is several Earth masses, and these objects are massive enough to migrate noticeably. 

For most objects just above the pebble isolation mass, the evolutionary tracks become nearly horizontal. These objects migrate slowly, but gas accretion rates are very low, especially for the low-mass planets inside 1 AU, so the masses hardly change. Gas accretion rates increase rapidly with planetary mass, while migration rates decrease with mass for the best-fit value of the migration parameter $x$. As a result, the evolutionary tracks curve upwards, becoming nearly vertical in some cases when planets undergo runaway gas accretion. Above about 100 Earth masses, the evolutionary tracks begin to flatten again. These objects are so massive that their gas accretion is limited by the inward flow of gas through the disk, which becomes quite small late in the disk's lifetime.

Figure 6 shows the evolutionary tracks for the same planets on a plot of mass versus time. The solid curves show objects accreting pebbles. The dashed curves show objects above the pebble isolation mass that can only grow by accreting gas. A wide range of pebble accretion rates is apparent, depending on the disk mass, time, and location of the planet. Only a small fraction of the planets grow massive enough (a few Earth masses) to undergo appreciable gas accretion. Gas accretion goes through several phases: (i) slow early stage when gas accretion is limited by the cooling timescale, (ii) rapid accretion when the cooling timescale becomes short, and (iii) later slowing of gas accretion as the supply of gas becomes limited by the disk accretion rate. 

Most of the gas giant planets begin rapid gas accretion after 1 My. Several objects do not start rapid gas accretion until close to the end of the simulation at 5 My. In some cases this is because the object didn't reach the pebble isolation mass until 1--2 My into the simulation. Other objects reached the pebble isolation mass at earlier times, but gas accretion was initially very slow due to the low mass of the planet at this point. Thus the number of giants that form depends on multiple parameters: the pebble isolation mass (and thus $\fiso$), the timescale required to reach this mass (which depends on $M_0$, $\tpeb$ and $\alpha$), and the rate of gas accretion (dependent on $\tcool$).

%
%
\section{The 4-Planet Case} \label{sec:n4}
\begin{figure}
\plotone{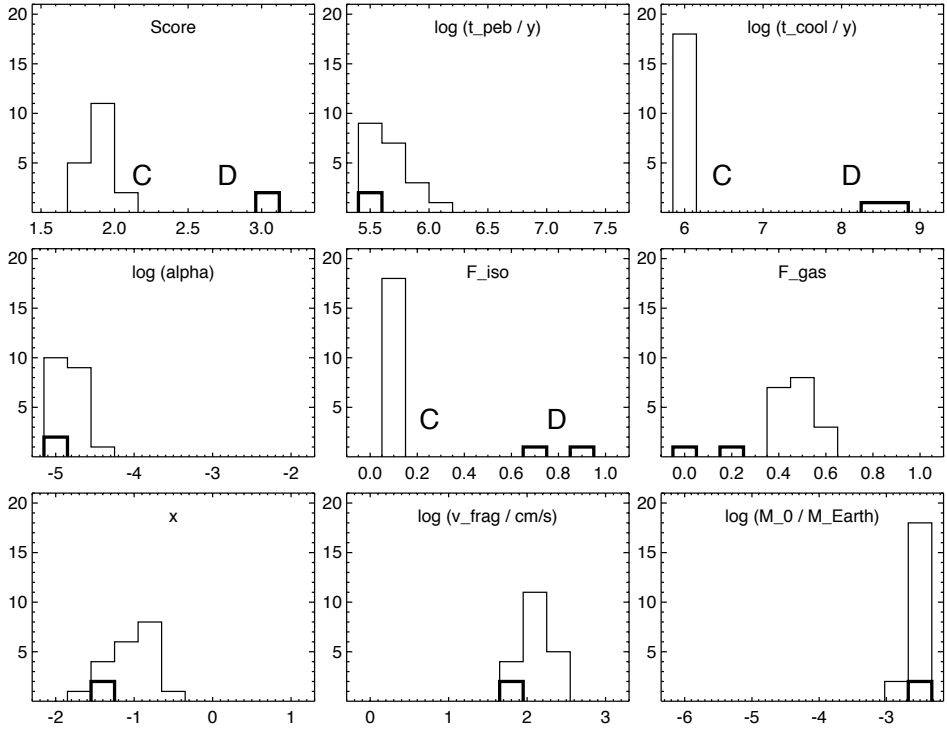}
\caption{Distribution of scores and best-fit model parameters for 20 optimization runs for systems with 4 planets per star. The best fit solutions are clustered into two groups labelled C and D.}
\end{figure}

In this section, we look at systems with 4 planets per star. In our model, multiple planets in the same system compete to accrete the available fluxes of pebbles and gas. Migrating planets with converging period ratios can also be trapped in the 2:1 mean-motion resonance, synchronizing their future orbital migration.

Figure~7 shows the final score for 20 runs of the optimization scheme using different random number seeds. The figure also shows the distributions for each of the 8 model parameters for the 20 runs. Eighteen of the 20 cases have converged to almost the same solution, labelled Solution C in the figure. The remaining 2 runs found a different solution with a substantially poorer fit (higher score) labelled Solution~D. 

Solution C for the 4-planet fit is an improvement over the single-planet Solution A. The mean scores are 1.90 and 2.38 respectively. Clearly, adding more planets per system allows a better fit to the observations. This makes intuitive sense in some cases. For example, \citet{foreman:2016} estimated that the Kepler data are consistent with 2 planets per star, which obviously can't be satisfied by single-planet systems.

Most of the parameter values are similar for solutions A and C (compare Figures~2 and 7). As before, the best fit solution in Figure~7 favors a large initial protoplanet mass, a low turbulence level, a short atmospheric cooling time, and a short pebble drift lifetime. A few differences are apparent. The values of $\fiso$ are smaller by about a factor of 2--3 in the 4-planet solution, so planets become isolated from the inward flux of pebbles at smaller masses than the single-planet case. In addition, the migration exponent $x$ is more negative, reducing migration rates for massive planets in the 4-planet systems compared  to the single-planet systems. However, the fact that most parameter values are similar in the  two cases suggests the main reason why the 4-planet fits are better is simply that more planets are present.

\begin{figure}
\plotone{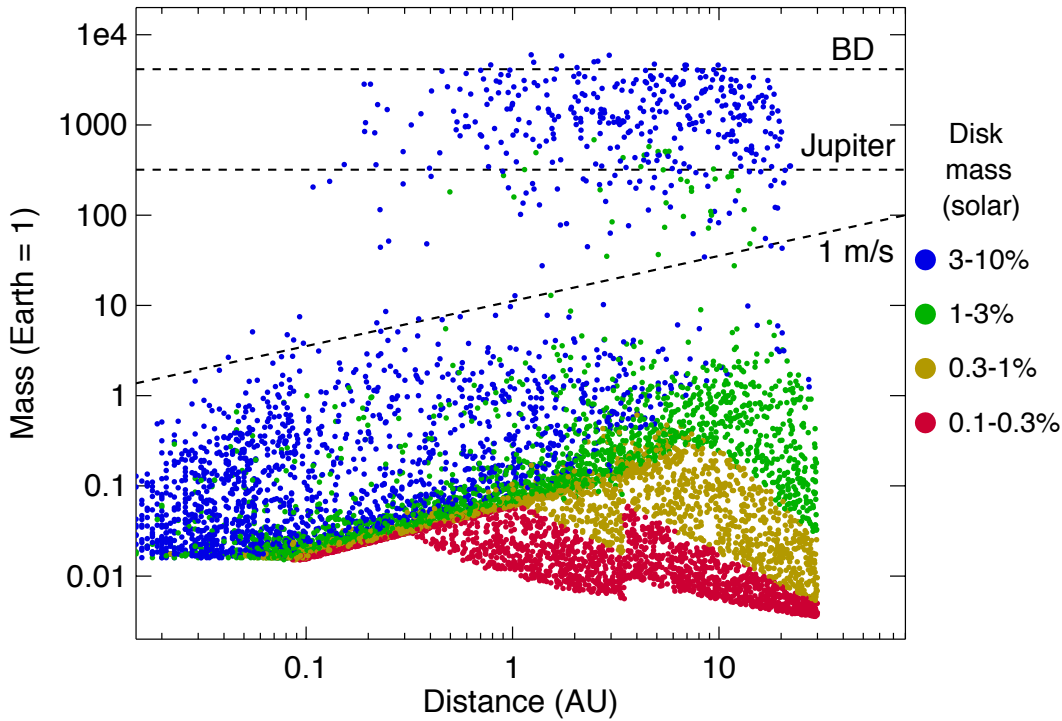}
\caption{Final masses and orbital distances of planets in 2000 systems, each with 4 planets, for the best fit model. The upper two dashed lines indicate 1 and 13 Jupiter masses (the latter labelled ``BD'' for brown dwarf). The lowest dashed line indicates a planetary mass that generates a 1 m/s Doppler radial velocity signal in the star. Different symbol colors indicate planets that formed in disks with different initial masses.}
\end{figure}

Solution D in Figure 7 is considerably poorer than Solution C, and also worse than the single-planet solutions in Figure~2. Systems formed using Solution~D rarely contain giant planets and never contain brown dwarfs. This is due to the long atmospheric cooling times (large $\tcool$), resulting in very slow gas accretion rates. The near failure to form giants means that Solution~D is only able to fit the observational constraints for low-mass planets. Solution~D does achieve a somewhat better fit for the low-mass planets than Solution~C (see Table 3), but this is not enough to offset the very poor results for giant planets and brown dwarfs. Hence solution~D achieves a poor score overall.

Table 3 shows the frequency of planets produced by two of the best-fit models compared  to the observations for the 31 regions of observational phase space described in Section~3. The table gives examples for Solutions~C and D, and indicates the discrepancy between model and observations in terms of the estimated error $\sigma$ on the observations. 

Comparing Tables~2 and 3, we see that the 4-planet Solution~C achieves overall improvements in all 7 groups of observational constraints (the seven subsections of Section 3) except for the brown dwarf frequency, which is already matched almost perfectly by Solution~A. Two notable improvements are (i) the number of planets per star  with small radii and $50<P<300$ days, and (ii) the number of planets per star with $2<P<25$ years.  The observational surveys suggest both these frequeuncies are large (0.77 and 2.00 planets per star), so it is not surprising that adding more planets per star improves these statistics the most. The smallest improvement between Solutions~A and C (other  than the brown dwarf frequency) is in the number of stars with planets more massive than Jupiter. This statistic improves only marginally.

\begin{figure}
\plotone{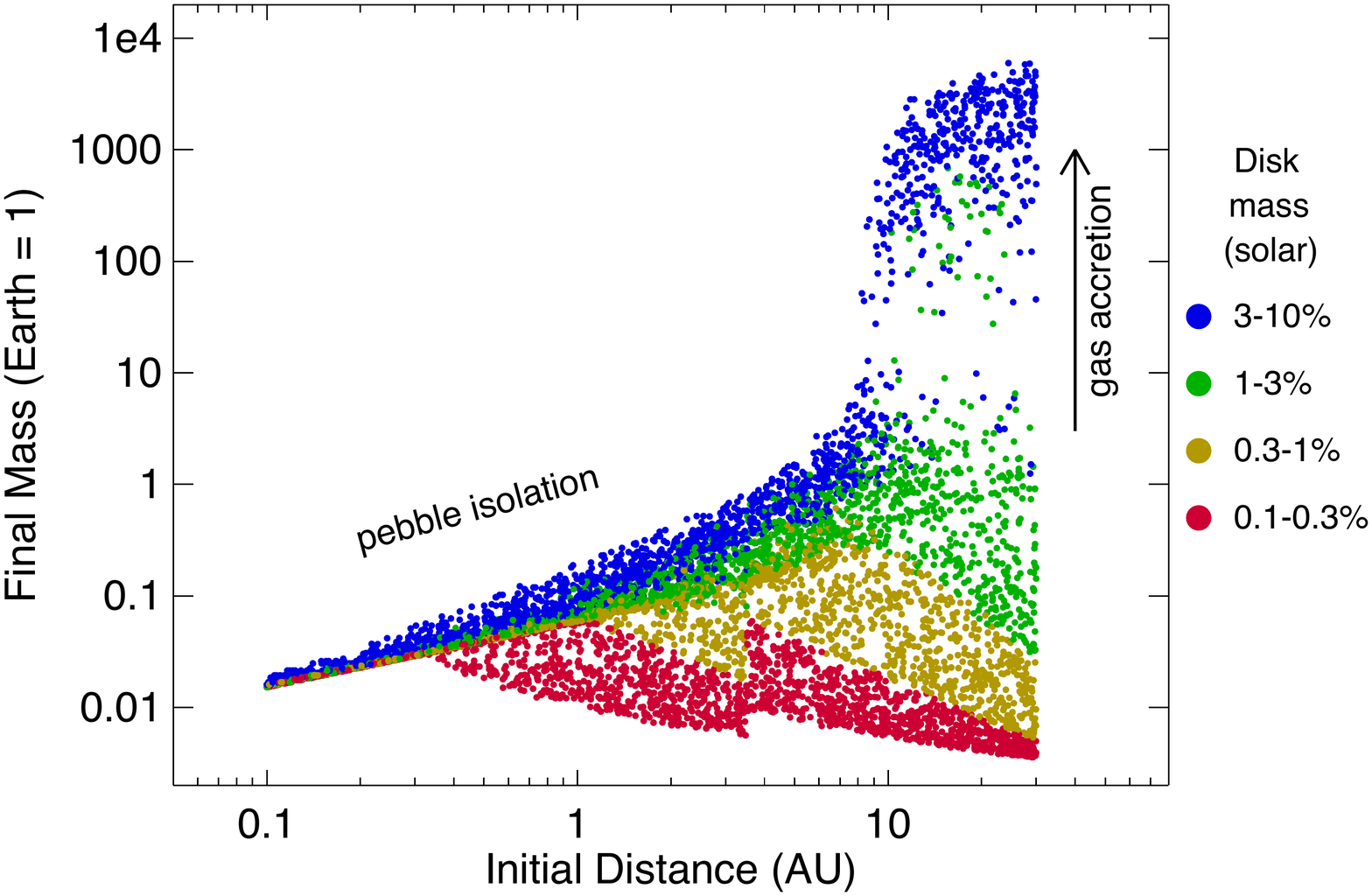}
\caption{Final planetary masses as a function of {\em initial\/} orbital distance for 2000 systems, each with 4 planets. Different symbol colors indicate planets that formed in disks with different initial masses. Planets initially inside 9 AU never grow above the pebble isolation mass, which is too low for the planet to undergo significant gas accretion. Planets that start beyond 9 AU in massive disks can become large enough to accrete gas and form giant planets.}
\end{figure}

For both Solutions A and C, the poorest fits to the observational data are for super Earths with periods less than 100 days. This is especially true for planets with radii between 2 and 3 Earth radii, where the model frequencies are typically 4 or 5 standard deviations below observations. This may mean that the model for gas accretion used here is flawed, either underestimating the formation of planets with modest gas fractions, or allowing these objects to grow into gas giants too frequently. A second possibility is that the pebble isolation mass depends on orbital distance in a more complicated way than we have assumed, in particular it could be enhanced specifically for objects in the inner disk. A final possibility is that yet more planets are required, especially in the inner disk.

Figure~8 shows the final orbits and masses for the planets for a Solution~C case. The figure contains 4 times as many points as Figure~3 since there are now 4 planets per star rather than 1. The distributions in the two figures have many features in common. Both feature a gap in the distribution for planets between about 10 and 50 Earth masses due to rapid gas accretion rates for these objects. Both figures exhibit a pile up of planets between 0.02 and 0.09 AU where migrating planets have reached the inner edge of the disk and migration ceases. There is also a strong correlation between disk mass and final planetary mass in both figures, with observable planets failing to form in low-mass disks in either case.

While the overall number of points is 4 times higher in Figure~8 compared to Figure~3, the number of giant planets has not increased by the same amount. In the single planet case, 230 planets formed with $M>30M_\oplus$, while in the 4-planet case, 415 such planets formed. This represents an increase of less than a factor of 2. Giant planet formation is clearly less efficient for the model parameters in Solution~C than Solution~A. Comparing planets more massive than Jupiter in the two figures, we find that a substantial fraction in the single-planet case formed in disks with masses 0.01--0.03$M_\odot$ (green symbols). In the 4-planet case in contrast, almost all planets larger than Jupiter formed in the most massive disks indicated by the blue symbols. The giant planets in Figure~8 are somewhat more concentrated towards larger distances than those in Figure~3 due to the more negative migration exponent $x$ and slower resulting migration rates.

\begin{figure}
\plotone{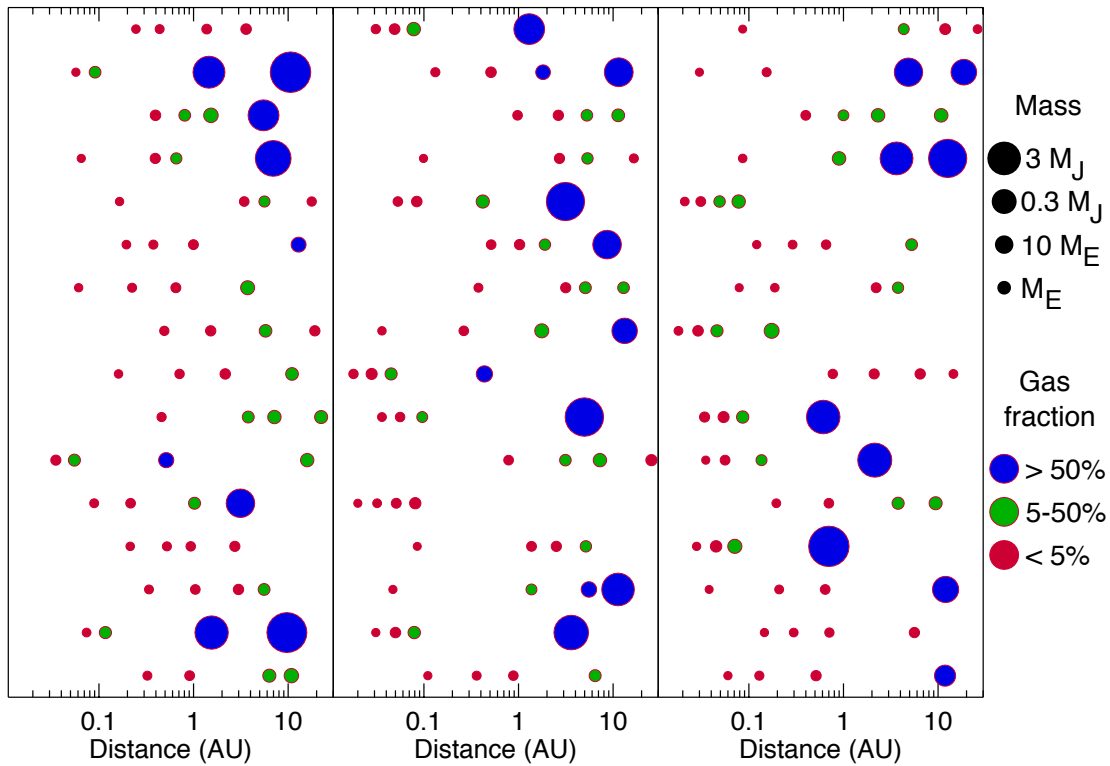}
\caption{Final planetary systems for 48 stars from an optimization run with 2000 stars with 4 planets per system. Each row of symbols shows the planets formed in a single system. Symbol size and color indicate the mass and gas mass fraction of the planet.}
\end{figure}

Figure 9 shows the final masses of the planets for Solution~C of the 4-planet case as a function of the {\em initial\/} orbital distance. This can be compared with Figure~4 for single-planet systems. Planets forming in low-mass disks (red and yellow points) occupy similar regions of phase space in the two figures. Almost all these planets are less massive than Earth. The distributions are also similar for planets forming in somewhat more massive disks (green points), at least for planets below about 3 Earth masses. The main difference between Figures~4 and 9 for low mass planets is that the pebble isolation mass at a given distance is lower in the 4-planet systems than single planet systems due to the smaller value of $\fiso$. 

The distribution of giant planets shows larger differences between Figures~4 and 9 than the low mass planets. The lower pebble isolation mass in the 4-planet case means that the distance at which planets can become massive enough to undergo significant gas accretion is greater. Thus, giants have to form further from the star in the 4-planet case, and fewer giants form overall. This improves the model fit to the observations by increasing the abundance of super-Earths  relative to giant planets. As noted earlier, the great majority of giant planets in Figure~9 form in the most massive disks (blue symbols). Giant planet formation in disks with masses less than 3\% of a solar mass is restricted to a narrow range of initial distances between 11 and 22 AU, while giants in more massive disks can form anywhere beyond 9 AU. Migration then transports these objects to a wide range of final distances, all the way to the inner edge of the disk in some cases.

Figure 10 shows a sample set of 48 final planetary systems for one of the Solution C runs for the 4-planet case. Each row of symbols shows the planets orbiting a single star. The size of each symbol indicates the final mass of the planet, while the symbol color shows the gas mass fraction of the planet. A wide variety of systems form in the model, but some common trends are apparent. In each system, planetary mass and gas mass fraction tend to increase with increasing distance from the star. The inner planets in each system tend to be small and have a low gas fractions. Some of the outer planets are similar but others are gas giants. Where gas giants form they are almost always one of the outer two planets in the system. Most of these features can be attributed directly or indirectly to the increase in pebble isolation mass with distance from the star.

\begin{figure}
\plotone{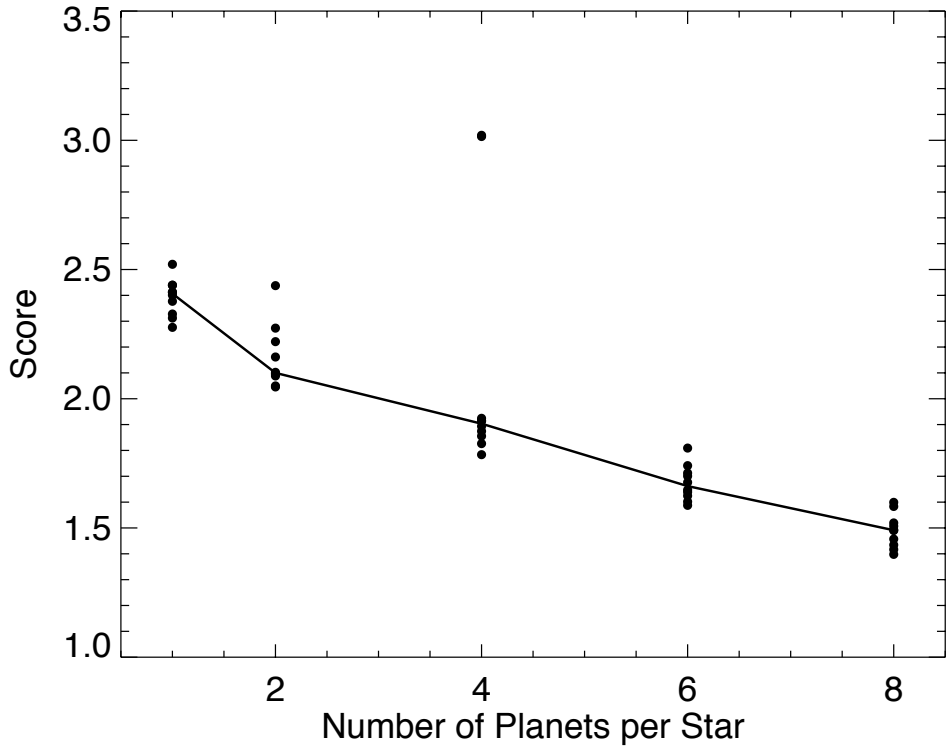}
\caption{Best-fit score versus number of planets per star $N$. Circular symbols show scores for 10 optimization runs for each value of $N$.  The line shows the median score for each $N$.}
\end{figure}

Quite a few systems contain chains of 2--4 planets in resonance with one another. These can be identified as the planets with the closest horizontal separation in the figure. (Note that three and four-planet resonant chains are equally spaced on the logarithmic scale of the horizontal axis of the figure.) Resonant chains usually involve the innermost planets, although there are exceptions. Interestingly, the chains of resonant inner planets are commonly accompanied by a giant planet orbiting further out. This appears to be because giants tend to form in massive disks, and these are also the disks that cause the highest migration rates for low-mass planets, increasing the chance of resonant capture.

So far all the cases we have considered contained either 1 or 4 planets per star. We also ran cases with other numbers of planets. Figure  11 shows how the best fit score varies with the number $N$ of planets per system. For each value of $N$, we ran 10 optimization runs with different random number seeds. The scores for each $N$ are shown by the symbols in Figure~11, while the line shows the median score for each $N$.

The median score decreases monotonically with increasing $N$. Thus continuing to add more planets per star tends to improve the fit with observations. The largest improvement occurs when going from the single-planet case to 2-planet systems, but we see a steady improvement beyond $N=2$ as more planets are added. For the 8-planet case, all the runs converged towards a similar solution with a mean score of 1.49 compared to 1.90 for Solution~C for the 4-planet case. The improvement arose largely due to a better fit to the observed frequency of low-mass planets, especially planets with $P<100$ days \citep{fulton:2017} and $50<P<300$ days \citep{burke:2015}. There was little corresponding improvement in the statistics for giant planets.

\begin{figure}
\plotone{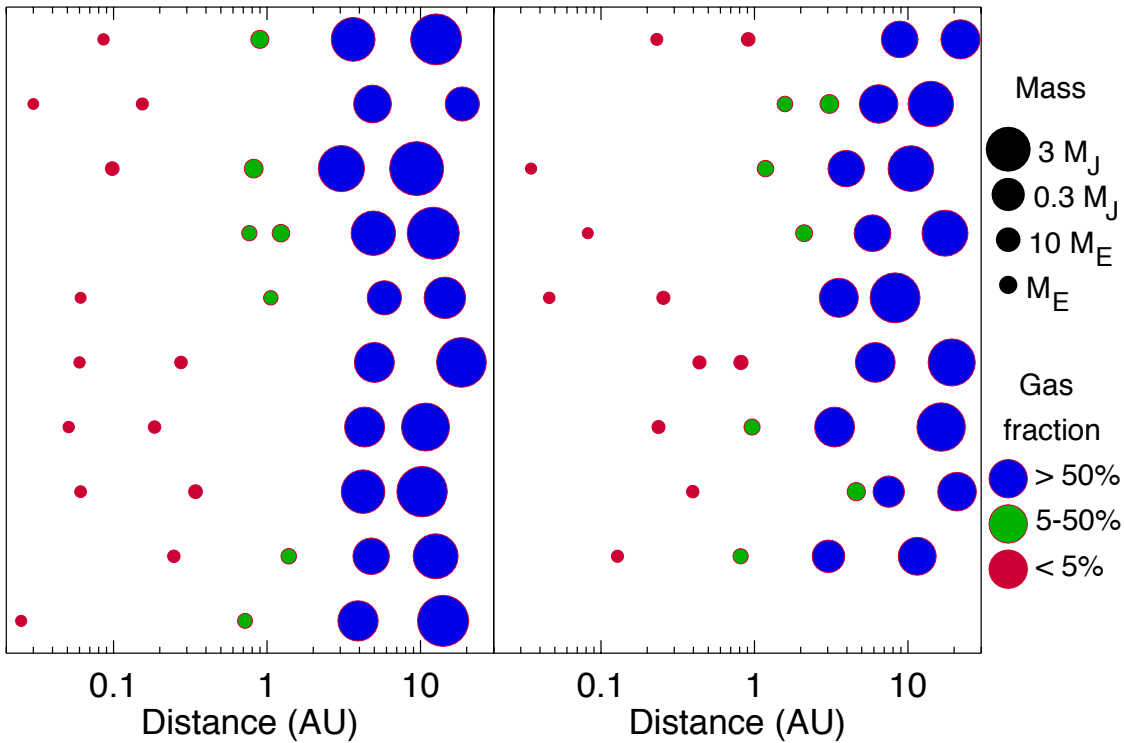}
\caption{Solar system analogs produced by the best-fit model for systems with 4 planets. Analogs are defined as systems with two planets larger than 50 Earth masses beyond 3 AU, and no planets more massive than 2 Earth masses inside 3 AU.}
\end{figure}

For all the cases considered here, we limited the number of planets to be no more than 8. Adding more planets typically violates the condition that the orbits are initially separated by more than their 2:1 mean motion resonance. It is plausible that using more than 8 planets could further improve the fit with observations. However, this would require a more complicated model than the one used here, including a more sophisticated treatment of planetary interactions and resonant capture.

%
%
\section{Solar System Analogs}
In this study we have specifically tried to match the observed distribution of extrasolar planets using our model for planet formation. The best fit solutions produce a wide variety of planetary systems (see Figure~10) as expected since the observed systems are also very diverse. This naturally raises the question of whether the model forms any systems like the Solar System even though this wasn't a goal of the fitting procedure. We address this question in this section.

The Solar System contains many objects, each with many characteristics, so the definition of a Solar System analog is subjective as a result. Here we adopt two simple criteria to identify analogs in the systems formed by the model, choosing criteria that are relevant to the kind of observations that can be made at present. We define a Solar System analog as a system with:
\begin{enumerate}
\item At least 2 planets with masses above 50 Earth masses and orbital distances $>3$ AU.
\item No planets interior to 3 AU with masses above 2 Earth masses.
\end{enumerate}

\begin{figure}
\plotone{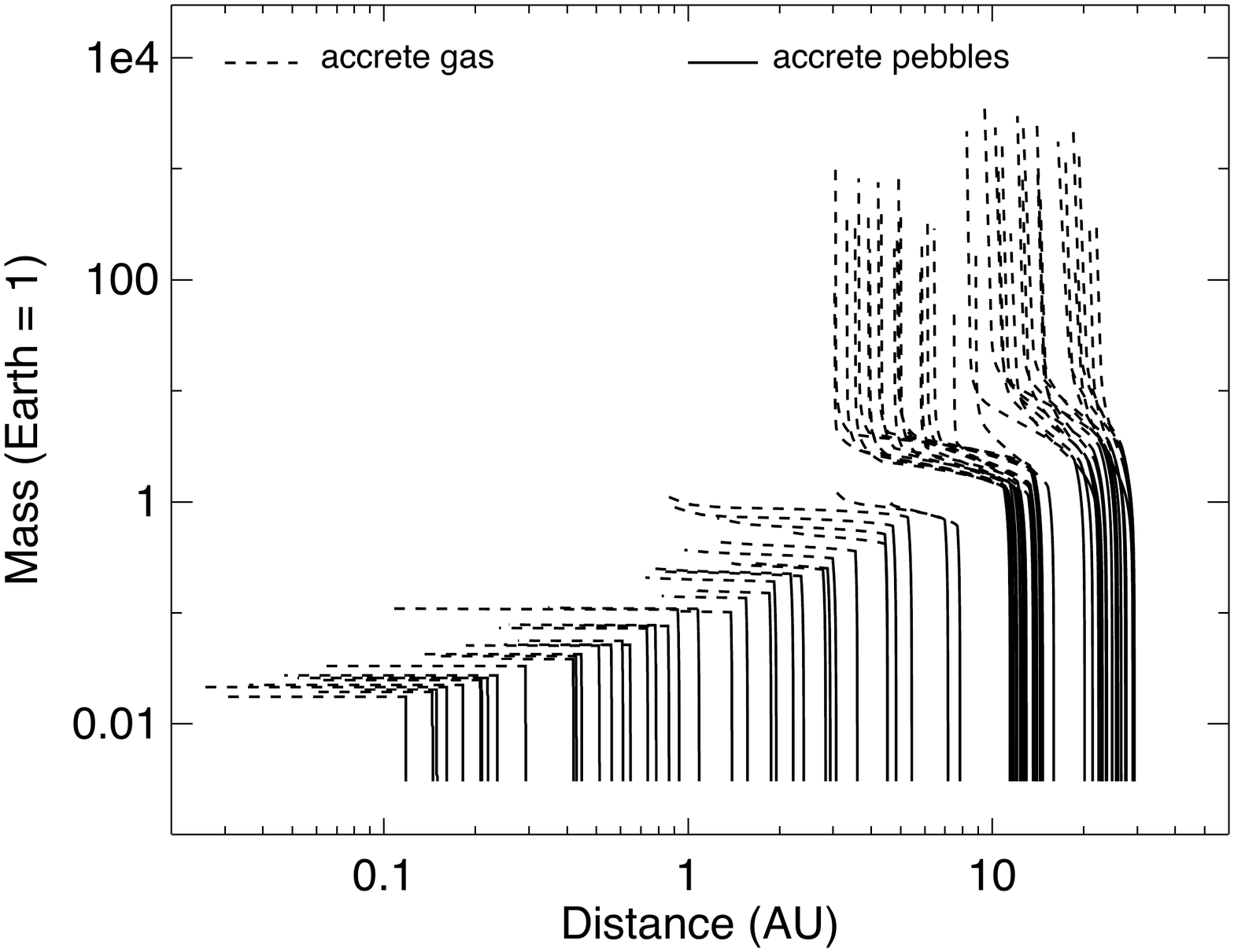}
\caption{Evolutionary tracks in mass-orbital distance space for the planets shown in Figure~12.}
\end{figure}

Figure~12 shows all systems identified as Solar System analogues according to these criteria for the 4-planet case shown in Figures~7--10. Out of 2000 stars, there are 19 Solar System analogs, that is 0.95\% of systems. If we consider all the 4-planet optimization runs that converged to Solution~C, the frequency of Solar System analogs is also about 1\%.  For runs with 6 planets the result is similar, with roughly 1\% of systems resembling the Solar System, while for runs with 8 planets, the frequency of analogs rises to about 2\%.

Clearly Solar System analogues are rare according to the fairly liberal criteria adopted here. They are even rarer if we adopt additional constraints such as the absence of planets interior to Mercury's orbit. This result needs to be interpreted with caution, however, since  the results presented here were optimized to fit the observed distribution of extrasolar planetary systems, and those found to date rarely resemble the Solar System. If future surveys find more systems with giants on long-period orbits, for example, then fitting to this updated distribution of planets is likely to increase the frequency of Solar System analogs. For now we simply note that a model that provides a reasonable fit to the observed extrasolar planets does occasionally form systems like our own.

Looking at the Solar System analogs for the 4-planet case, we find that they all have some characteristics in common. These systems form in massive disks, containing at least 0.03 solar masses of material. The disks also have large radii, at least 100 AU in 18 of the 19 cases in Figure~12. Thus, these systems contain a large amount of solid material to build the solid cores of giant planets. At the same time, the gas surface density is relatively low due to the large disk radius, reducing the effects of migration (see Figure~13). Finally, the outer two planets begin at large distances (greater than 10 and 20 AU respectively). This has two advantages. Firstly, the pebble isolation mass is large at these distances, allowing substantial gas accretion to occur. Secondly, giant-planet cores tend to reach the pebble isolation relatively late, when the gas surface density has declined substantially from its initial value, again limiting the degree of migration and preventing giants from entering the inner disk.

%
%
\section{Discussion}
In this paper we have developed a simple model for planet formation with 8 free parameters that describe some of the main uncertainties in the physics of protoplanetary disk evolution, planetary growth, and planet orbital migration. We ran this model for sets of 2000 stars with protoplanetary disk masses and radii compatible with observed distributions, and random initial protoplanet orbital distances. The planets produced by the model were compared with 7 published studies that have estimated the orbital and mass/radius distributions of extrasolar planets and brown dwarfs. The model parameters were then improved iteratively using a particle swarm optimization scheme in order to find the best-fit set of parameters.

The planet formation model considers a small number of protoplanets embedded in an evolving protoplanetary disk. Planets accrete solid mass in the form of pebbles that drift inwards through the disk due to gas drag. Objects that reach a certain size---the pebble isolation mass---perturb nearby gas sufficiently to halt the inward flux of pebbles. Planets above this mass can accrete gas from the disk at a rate that depends on the planetary mass, the cooling timescale for the planet's atmosphere, and the inward flux of gas. Low-mass planets undergo inward type-I migration due to tidal interactions with the disk. Planets massive enough to open a gap in the disk inwards migrate more slowly at a parameterized rate. Converging migrating planets can be trapped in their 2:1 mean-motion resonance.

For both single-planet and multi-planet systems, the closest fit to the observations occurs for large initial protoplanet masses, low disk viscosities, short pebble drift lifetimes, and short cooling times for planetary atmospheres when gas accretion begins. The first three of these promote rapid growth of solid planets, while the last promotes rapid gas accretion. All four of these parameter values tend to lie close to the maximum or minimum permitted values. The best fit solutions find that planets reach the pebble isolation mass at a small fraction of the mass needed to open a gap in the gas disk. The best fits also indicate that pebbles undergo collisional disruption at speeds around 1 m/s, and that gas giants can accrete at most about half of the inward flux of gas through the disk.

Overall, the best-fit models do a good job of matching the observed frequency and orbital distribution of giant planets and the observed abundance of brown dwarfs. The best fits provide a poorer match for the observed population of super Earths on short-period orbits. This is especially true for objects with radii 2--3 times that of Earth, which are substantially under represented in the model compared to observations. A few fits provide a better match to the observed super Earth population but at the expense of greatly underestimating the number of giant planets and brown dwarfs.

We find that adding more protoplanets per system in the model improves the fit substantially compared to single-planet systems. This is significant because single-planet systems are often used in population synthesis studies. The best fit solutions continue to improve as more planets are added up to 8 per system, which is the maximum number that matches our constraint that neighboring objects begin exterior to their 2:1 mean-motion resonance. The main effect of adding more protoplanets is to increase the number of super Earths that form relative to the number of giant planets, which tends to improve the fit to the observed planetary distribution.

Pebble isolation appears to be the key process that determines the characteristics of planetary systems produced by the model used here. Typically, planets do not start accreting gas until they reach the pebble isolation mass, since heat released by infalling pebbles prevents the atmosphere from cooling and contracting so that gas can flow inwards \citep{lambrechts:2014}. In our model, the pebble isolation mass increases with distance from the star since it is proportional to the mass needed to open a gap in the gas disk, which also increases with distance \citep{rafikov:2002}. Protoplanets also tend to grow faster closer to the star because pebble accretion is more effective.

A combination of these factors determines where gas giant planets can form. In the inner disk, most planets quickly grow to the pebble isolation mass. In principle, gas accretion can begin at this point. However, the planetary masses are small, so gas accretion is extremely slow. These planets never accrete more than a small amount of gas as a result. In the outer disk, the pebble isolation mass is much larger. Any planets that reach this size can undergo rapid gas accretion and become giant planets. However, in most disks, the growth of solid planets in the outer disk is too slow to reach the pebble isolation mass before the disk disperses, especially since the pebble flux dwindles rapidly at later times. 

Only when the disk mass is large can solid planets in the outer disk grow fast enough to reach pebble isolation and undergo rapid gas accretion to form giant planets. Giants migrate inwards during and after their formation, although the degree of migration varies depending on the planetary mass, the time, and the nature of the disk. Thus, the final orbits of giant planets tend to occupy a wide range of distances extending to the inner edge of the disk.

The solutions favored by the optimization scheme lead to rapid early growth of solid planets due especially to the small values of the turbulence strength $\alpha$ and the pebble flux decay timescale. Partly this is because rapid growth increases the chance that giant planet cores can form early enough to undergo the prolonged stage of slow gas accretion before runaway gas accretion begins. At least as important, however, is the fact that the gas surface density declines over time, and this means that the pebble isolation mass decreases as well. In order to form a giant planet, a protoplanet must exist in a part of this disk where the pebble isolation mass is large enough for runaway gas accretion to occur, and the object must grow fast enough to reach the pebble isolation mass before it has decreased too far.

In this study, we find very low values for the pebble isolation mass. For a typical gas surface density of 100 g/cm$^2$ at 5 AU, the isolation mass is around an Earth mass for the best-fit single-planet case, and 2--3 times smaller for runs with 4 or more planets. By contrast, \citet{lambrechts:2014} found a pebble isolation mass of roughly 20 Earth masses at 5 AU. We note however that these authors used a high disk viscosity with $\alpha=0.006$ compared to $\alpha\sim10^{-5}$ for our best fit models. For viscosities this low, analytic calculations and hydrodynamical simulations both suggest a planet with a mass of about 10 Earth masses would open a deep gap at 5 AU \citep{rafikov:2002, duffell:2013}. The planetary mass needed perturb the gas enough to stop inward drifting pebbles is likely to be at least a few times smaller than the gap-opening mass \citep{lambrechts:2014}. Thus, the very small pebble isolation masses found here may not be too unreasonable.

In the planet formation model used here, solid planets grow entirely due to pebble accretion rather than by accreting planetesimals or merging with other planets. Runaway and oligarchic growth of planetary embryos, which feature prominently in the standard picture of planet formation \citep{wetherill:1993, kokubo:1998} do not occur here. One consequence is that terrestrial planets form rapidly, within the lifetime of the gas disk. Earth is thought to have formed in $\sim 10^8$ years \citep{jacobson:2014}, much longer than the lifetime of a typical disk \citep{haisch:2001}, with much of its growth involving occasional giant impacts between planets after the gas dispersed. This discrepancy can be reconciled by treating the small, rocky planets that form in our simulations as tracers of a larger number of planetary embryos present in the inner disk that will eventually merge into a handful of terrestrial planets. We note, however, that many short-period super Earths in extrasolar systems appear to contain significant amounts of gas, so these objects must have formed while the gas disk was still present.

An additional potential problem with forming Earth or its precursors from pebbles is that most of these pebbles probably originated in the outer solar nebula and drifted inwards to Earth's location. Initially at least these pebbles would be volatile rich since they formed in a cold environment. Most of these volatile constituents would be lost as pebbles entered the hotter inner nebula. However, it is unclear whether the surviving components can be reconciled with Earth's current composition cosmochemically.

The planet formation model used here was designed to be extremely efficient so that it could be run for the large number of times required by the optimization scheme. Some potentially important physical processes were simplified or neglected as a result. For example, we adopt a relatively simple model for type-I migration appropriate for tidal torques due to gas in a non-magnetized, isothermal disk \citep{tanaka:2002}. More detailed models for migration have been developed which could change the conclusions of this paper \citep{paardekooper:2011, baruteau:2011, benitez:2018}. We  also have not considered planetary orbital evolution after the protoplanetary disk disperses due to planet-planet scattering, planetesimal driven migration, or stellar tides.

In our model, the radial temperature profile varies as $1/\surd a$ where $a$ is orbital distance. This is roughly the thermal profile for a disk that is heated entirely by the central star \citep{chiang:1997}. However, temperatures in the inner part of the disk may be dominated by heat released by viscous accretion, which typically leads to a steeper radial temperature profile \citep{chambers:2009}. This would alter several parts of the planet formation model used here such as the pebble capture rate, the planetary gas accretion rate, and the migration of massive planets. Perhaps most importantly, changing the thermal profile would alter the aspect ratio of the inner disk and thus the pebble isolation mass.

To test whether this effect is important, we carried out sets of 20 optimization runs for 1- and 4-planet systems using a modified temperature profile. Here, temperature varies as $1/\surd a$ outside 1 AU, and as $1/a$ inside 1 AU, normalized to 300 K at the transition point. For the 1-planet case, the new best-fit solution is poorer for the modified disk thermal proflie. However, the best-fit solution is somewhat improved for the 4-planet case with a score of 1.68 for the best optimization run compared to 1.79 previously. This improvement is almost entirely due to increased production of super-Earths (radii 1--4 Earth radii) with periods of less than 100 days---the region considered by \citet{fulton:2017}. The frequency of these planets increases by roughly 60\% for the best-fit case, due in part to a modest increase in the pebble isolation mass. This suggests future population synthesis studies may need to consider carefully the thermal structure of the disk, as done for example by \citet{ndugu:2018}.

Pebbles in the simulations described here typically have Stokes numbers around 0.01 (sizes of order 1 cm). It is unclear whether these pebbles are large enough to form protoplanets via the streaming instability, although some recent simulations suggest this is possible \citep{yang:2017}. It is also uncertain whether protoplanets would be as large as the Pluto-sized objects favored by the optimization scheme. However, only a handful of such bodies need to form, so these may represent the high mass tail of the distribution seen in numerical simulations \citep{schafer:2017} .

The best fit for any of the optimization runs is for a case with 8 planets with a score of 1.4. This solution lies 1.4 standard deviations from the estimated observational constraints on average. Other solutions are poorer, in particular when fewer planets are included. This mismatch could be due to some of the shortcomings of the model discussed above. It is also possible that the different sets of observational constraints used here are not mutually compatible within errors. Some of the published observational constraints that we use only provide statistical errors. However, \citet{burke:2015} found that there are probably systematic errors of at least comparable magnitude. If this is true of the other observational surveys it would improve our model fit, although the details of the fit could change as a result.

In this study, we considered ensembles of 2000 stars with different protoplanetary disk properties and initial protoplanet orbits. Typically, a different ensemble of stars was used for each optimization run. Different runs sometimes found different best-fit model parameters, implying that an ensemble size of 2000 is small enough that the random, discrete nature of the initial conditions can influence the result. This is particularly true for the random initial orbits of the planets, since the outcome is quite sensitive to where planets are located.

We could have considered larger ensembles. However, given that we currently have observations of only a few thousand extrasolar systems, using a larger ensemble of synthetic systems probably won't tell us much. If the planet formation model used here is a reasonable reflection of reality, the results suggest some features of the observed extrasolar planet population are affected by stochastic events and small-number statistics. Interpreting these features will require a better understanding of the earliest stages of planet formation, especially which disk locations are most likely to generate protoplanets.

Despite this uncertainty, the optimization runs do yield some robust results. For example, the observations are best explained if protoplanets are initially large, have short atmospheric cooling timescales, and form in disks with low turbulence levels. This indicates that population synthesis studies can be a useful tool for constraining at least some aspects of planet formation.

%
%
\section{Summary}
We have developed a simple model for the growth and orbital evolution of a small number of protoplanets in an evolving protoplanetary disk. Planets accrete pebbles that drift inwards through the disk due to gas drag. Planets that cease accreting pebbles can accrete gas from the disk. Planets undergo inward migration, and can be captured in mean-motion resonances with one another. The model has 8 free parameters that quantify some of the main uncertainties in the physics of planet formation. We ran the model for sets of 2000 stars with protoplanetary disk masses and radii drawn from random distributions compatible with observations. The final planetary systems were compared with the observed extrasolar planet orbital and mass/radius distributions from seven published studies. Using this comparison, the model parameters were improved iteratively using a particle-swarm optimization scheme.

Our main findings from this study are
\begin{enumerate}
\item The characteristics of planetary systems when the disk disperses are mainly controlled by the pebble isolation mass. This is the mass of a planet that perturbs the local gas profile sufficiently to halt the inward flux of pebbles so that solid growth ceases. We find that the pebble isolation mass is small (less than 1 Earth mass) in the inner disk, but increases with distance. The pebble isolation mass decreases with time since it is related to the gas surface density.

\item Planets in the inner few AU of the disk generally reach the pebble isolation mass within the disk lifetime. However, these objects are too small to accrete appreciable amounts of gas, or migrate significantly. These form a population of sub-Earth mass planets, most of which are undetectable currently.

\item Planets in the outer disk accrete pebbles more slowly. In low mass disks, they never reach the pebble isolation mass, and do not accrete gas. In disks with masses $>0.01M_\odot$, planets in the outer disk can reach the pebble isolation mass within the disk lifetime, and some are massive enough to accrete large amounts of gas. These planets undergo varying degrees of inward migration, and form a population of gas giant planets with a wide range of final orbital distances.

\item The best model fits favor large initial protoplanet masses, short pebble drift lifetimes, and low disk viscosities. All these promote rapid early growth of solid planets. This has two advantages for forming gas giants: (i) objects reach the pebble isolation mass early while this mass is still large, and (ii) this allows sufficient time for the slow early stages of gas accretion to take place before the disk disperses. The best fits also favor short atmospheric cooling times (low opacities) allowing faster gas accretion.

\item The model typically does a good job of matching the observed mass and orbital distribution of extrasolar giant planets, and the observed abundance of brown dwarfs. The best fits do a poorer job of matching the distribution of super Earths, especially short-period planets with radii 2--3 times that of Earth.

\item Adding more protoplanets substantially improves the fit to observations. In particular, the match with the observed distribution of super Earths steadily improves as more planets per star are considered.

\item The best fit model solutions adequately reproduce the observed extrasolar planet population without requiring a reduction in (type-I) migration for low-mass planets. However, matching observations requires that migration rates decline substantially relative to type-I migration for massive planets.

\item Although this study was designed to match the properties of extrasolar systems only, we find that 1--2\% of systems with 4 or more planets can be considered reasonable analogs of the Solar System. These systems have two planets greater than 50 Earth masses orbiting beyond 3 AU, while all planets interior to 3 AU have masses less than 2 Earth masses.

\end{enumerate}

\acknowledgments
I would like to thank the reviewer, Michiel Lambrechts, for helpful comments and suggestions that improved this work.

%
%
\appendix
\section{Gas Accretion Rate}
In this appendix we derive an approximate expression for the gas accretion rate onto a planet embedded in a protoplanetary disk. We assume that the atmospheric luminosity is dominated by the energy released by infalling gas. We consider a two-layer model for the planetary atmosphere: an inner convective region and an outer radiative region. Quantities at the radiative-convective boundary (RCB) have subscript $b$; quantities at the outer edge of the atmosphere have subscript 0; and quantities at the core have subscript $c$.

We assume that the mass of the radiative region is small compared to the core, and that the gas here is ideal, so that the density $\rho$, pressure $P$ and temperature $T$ are related by
\begin{equation}
\rho=\frac{P\mu m_H}{kT}
\end{equation}
where $\mu$ is the mean molecular weight, and $m_H$ is the mass of a hydrogen atom.

The temperature and pressure in this region vary with radius $r$ as
\begin{eqnarray}
\frac{dP}{dr}&=&-\frac{G\mplan\mu m_H}{k}\frac{P}{r^2T}
=-r_B\frac{T_0}{T}\frac{P}{r^2}
\nonumber \\
\frac{dT}{dr}&=&-\frac{3\kappa L\mu m_H}{64\pi\stefan k}
\frac{P}{r^2T^4}
\end{eqnarray}
where $L$ is the atmospheric luminosity, $\kappa$ is the opacity, $k$ is Boltzmann's constant, $\stefan$ is the Stefan-Boltzmann constant, $\mplan$ is the total planetary mass (core plus atmosphere), and $r_B$ is the Bondi radius of the planet given by
\begin{equation}
r_B=\frac{G\mplan}{c_s^2}
\end{equation}

Thus
\begin{equation}
\frac{dT}{dP}=\frac{Q}{T^3}
\end{equation}
where
\begin{equation}
Q=\frac{3\kappa L}{64\pi\stefan G\mplan}
\end{equation}

The relation between $P$ and $T$ in  the radiative region is 
\begin{equation}
T^4=T_0^4+4Q(P-P_0)
\end{equation}

At the RCB, the logarithmic $T$-$P$ gradient matches the adiabatic gradient $\ad$, so that
\begin{equation}
QP_b=\ad T_b^4
\end{equation}

Assuming that $P_b\gg P_0$, we get
\begin{eqnarray}
T_b^4&=&T_0^4\left(\frac{1}{1-4\ad}\right)
\nonumber \\
QP_b&=&T_0^4\left(\frac{\ad}{1-4\ad}\right)
\end{eqnarray}

We will also need to calculate $r_b$. To do this, we note that temperature changes much more slowly than pressure in the radiative region, so to a first approximation we take
\begin{equation}
\frac{T}{T_0}=\left[1+\frac{4Q(P-P_0)}{T_0^4}\right]^{1/4}
\simeq
1+\frac{QP}{T_0^4}
\end{equation}

Thus
\begin{equation}
\int\left[1+\frac{QP}{T_0^4}\right]\frac{dP}{P}
\simeq
-r_B\int\frac{dr}{r^2}
\end{equation}

This gives
\begin{equation}
\log\frac{P_b}{P_0}+\frac{\ad}{1-4\ad}
\simeq\frac{r_B}{r_b}
\end{equation}
where we have assumed that $r_b\ll r_0$. The logarithmic dependence on pressure means that to a rough approximation we have
\begin{equation}
r_b\propto r_B\propto\frac{\mplan}{T_0}
\end{equation}
where the second relation follows from the definition of the Bondi radius.

In the convective region, we do not assume the gas is ideal, nor that the mass is small compared to the core. Instead, we assume that the density can be expressed approximately as a power law, with most of the mass concentrated near the core:
\begin{equation}
\rho=\rho_b\left(\frac{r_b}{r}\right)^n
\hspace{20mm} n>3
\end{equation}

The total mass of the atmosphere is given by
\begin{equation}
\matmos\simeq\int_c^b 4\pi r^2\rho dr
\simeq \frac{4\pi\rho_b}{(n-3)} r_b^nr_c^{3-n}
\end{equation}

Following \citet{lee:2015}, we assume that the energy liberated by infalling gas is approximately balanced by the luminosity. Since most of the atmospheric mass is concentrated near the core, this means that
\begin{equation}
\frac{d\matmos}{dt}\simeq
\frac{Lr_c}{G\mcore}
\end{equation}

Substituting for $L$ and $Q$ we get
\begin{equation}
\frac{d\matmos}{dt}\simeq
\left(\frac{64\pi\stefan\ad}{3\kappa}\right)
\frac{\mplan r_cT_b^4}{\mcore P_b}
\end{equation}

We also substitute for $P_b$ by noting that the gas is assumed to be ideal at this point, and using the relation between $\rho_b$ and $\matmos$:
\begin{equation}
\frac{d\matmos}{dt}\simeq
\left[
\frac{256\pi^2\stefan\mu  m_H\ad}
{3(n-3)k\kappa}
\right]
\frac{\mplan T_b^3r_c^{4-n}r_b^n}{\mcore\matmos}
\end{equation}

We now examine how the gas accretion rate depends on important quantities for two cases in which the core radius varies as the one third or one fourth power of the core mass, where the latter allows for some degree of gravitational compression of the core:
\begin{eqnarray}
\frac{d\matmos}{dt}
&\propto&
\frac{\mplan^{n+1} \mcore^{1/3-n/3}}
{\matmos\kappa T_0^{n-3}}
\hspace{22mm}r_c\propto \mcore^{1/3} \nonumber \\
&\propto&
\frac{\mplan^{n+1}}{\mcore^{n/4}\matmos\kappa T_0^{n-3}}
\hspace{20mm}r_c\propto \mcore^{1/4}
\end{eqnarray}

The value of $n$ is uncertain. Here, we consider two plausible values: $n=3.5$ and 4. Taking $n=3.5$, we get
\begin{eqnarray}
\frac{d\matmos}{dt}&\propto&
\frac{\mplan^{9/2} \mcore^{-5/6}}
{\matmos\kappa T_0^{1/2}}
\hspace{20mm}r_c\propto \mcore^{1/3} \nonumber \\
&\propto&
\frac{\mplan^{9/2}\mcore^{-7/8}}{\matmos\kappa T_0^{1/2}}
\hspace{20mm}r_c\propto \mcore^{1/4}
\end{eqnarray}

Taking $n=4$, we get
\begin{eqnarray}
\frac{d\matmos}{dt}&\propto&
\frac{\mplan^5\mcore^{-1}}
{\matmos\kappa T_0}
\hspace{20mm}r_c\propto \mcore^{1/3} \nonumber \\
&\propto&
\frac{\mplan^5\mcore^{-1}}{\matmos\kappa T_0}
\hspace{20mm}r_c\propto \mcore^{1/4}
\end{eqnarray}

\begin{figure}
\plotone{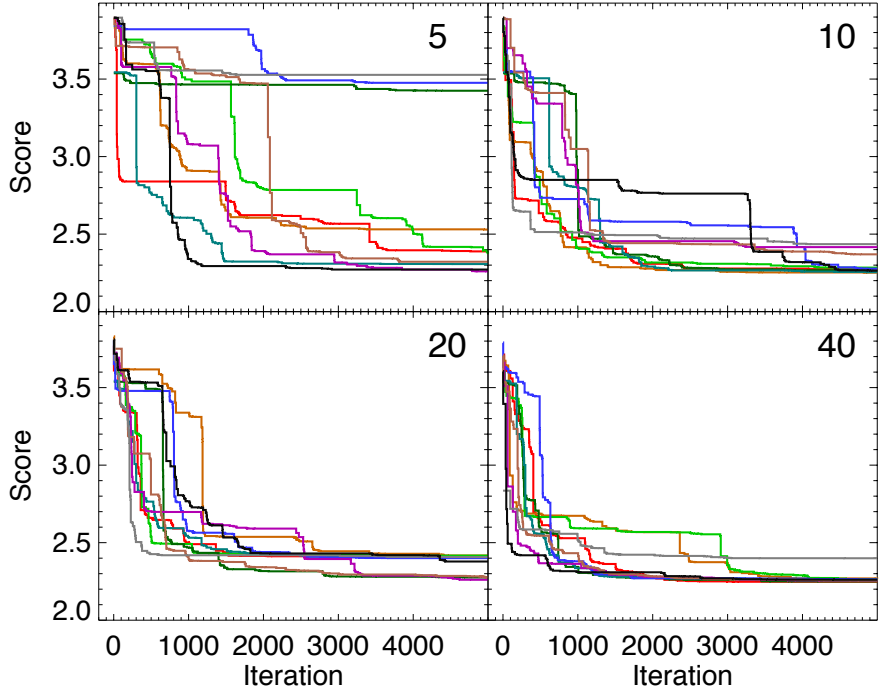}
\caption{Ten particle swarm optimization (PSO) runs with different initial PSO particle positions and velocities, but the same set of protoplanetary disk parameters and initial protoplanet orbits. Each run considers 2000 stars with a single planet, and is indicated by a different colored curve. The numbers in the upper right corner of each panel indicate the number of PSO particles used per run.}
\end{figure}

If we assume that $\mplan\sim \mcore$, and $n=3.5$ and $r_c\propto M_c^{1/3}$, we get the same dependence on $\mcore$ and $\matmos$ and $T_0$ and $\kappa$ as \citet{bitsch:2015} based on numerical calculations by \citet{piso:2014}.

In this study we adopt a slightly different expression suitable for both $\matmos<\mcore$ and $\matmos>\mcore$:
\begin{equation}
\frac{d\matmos}{dt}\propto\frac{\mplan^{9/2}}{\mcore\matmos T_0^{1/2}}
\end{equation}
where the constant of proportionality includes uncertainties in the opacity and adiabatic gradient.  Integrating this gives a gas mass that increases with the square root of time, similar to the $t^{0.4}$ dependence found by \citet{lee:2015}. During runaway growth, when the core mass is small compared to the gas mass, we get a growth rate of $d\mplan/dt\propto \mplan^{3.5}$ which is similar to the expression used by \citet{ida:2004} but with an exponent of 3.5 instead of 4.

%
%
\section{Particle Swarm Optimization Details}
In this appendix, we provide more details of the particle-swarm optimization (PSO) scheme and describe some convergence tests. Much of the history and theory behind the PSO scheme is described in an excellent review by \citet{poli:2007}. However, as these authors note, the use of PSO is still in its infancy, and the best way to use the scheme is still being established. 

In particular, the optimal values of the parameters used by the scheme to modify the particle trajectories ($C_1$ through $C_3$ in Eqn.~\ref{eq-pso}) may depend on the problem at hand. Small values lead to rapid convergence to a single solution but many miss other minima representing better solutions. Large values reduce this risk but allow particles to spend much of their time exploring parts of parameter space outside the region of interest. \citet{poli:2007} recommend using $C_1=0.7298$ and $C_2=C_3=1.49618$ but we find this typically leads to premature convergence  for our problem. After some trial and error, we find that $C_1=0.83$ and $C_2=C_3=1.65$ provide a good compromise between efficiency and ability to explore parameter space in detail.

\begin{figure}
\plotone{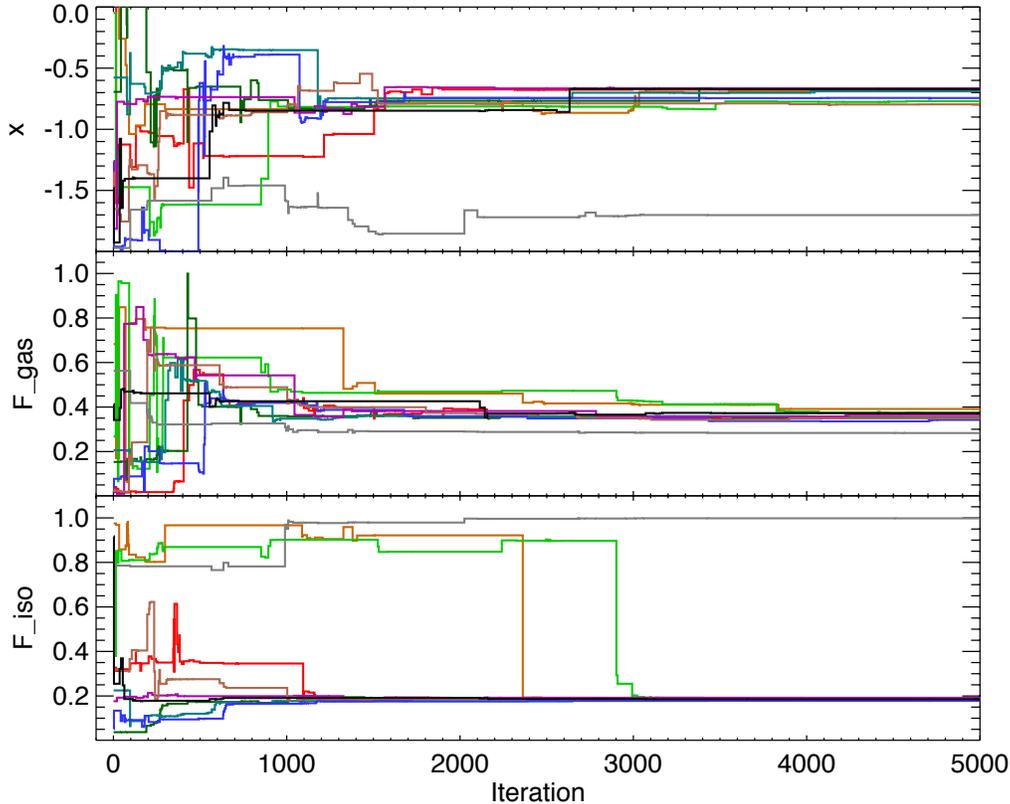}
\caption{Ten particle swarm optimization (PSO) runs with different initial PSO particle positions and velocities, but the same set of protoplanetary disk parameters and initial protoplanet orbits. (These are the same runs shown in the lower right panel of Figure~14). Each run considers 2000 stars with a single planet, and uses 40 PSO particles per run. The panels show the evolution of the best-fit values for 3 model parameters versus PSO iteration number.}
\end{figure}

The number of PSO particles is another free parameter. Using too few particles means that parameter space is poorly sampled, while using too many is inefficient since many particles can end up exploring the same region. To test the effects of this, we ran sets of 10 cases with different initial positions and velocities for the PSO particles, with each set using a different number of particles.  Each run considered $\nsys=2000$ stars, each with a single planet and randomly chosen disk properties. However, the set of initial disk parameters and initial protoplanetary orbits were the same for each of the 10 PSO runs, so the runs should in principle converge to the same solution.

Figure~14 shows the scores achieved versus iteration number for PSO runs using 5, 10, 20 and 40 particles respectively. For runs with 5 particles, the rate of convergence is rather poor. After 5000 iterations, the 10 runs fall into two groups, each converging  on a  different solution, with several runs in each group. For runs with 10 and 20 particles, the rate of convergence is better, but multiple runs end up converging  to different solutions. For 40-particle runs, the rate of convergence is slightly improved again. More importantly, 9 of the 10 runs converge to a single solution. In this case, there is little change after about 3000 iterations. For this reason we use 40 particles and run for 5000 iterations in the simulations described in the main text. For the single-planet case, each PSO run takes several hours of CPU time for the planet formation model used here, while multi-planet runs typically require a day or less.

\begin{figure}
\plotone{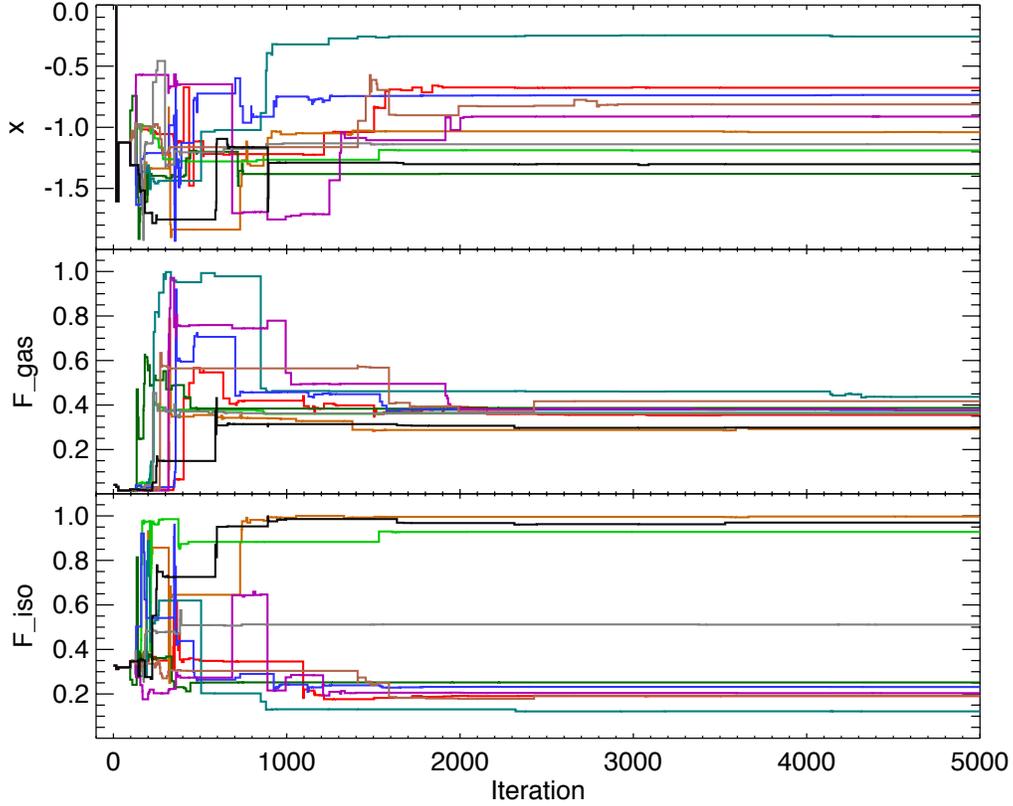}
\caption{Ten particle swarm optimization (PSO) runs with identical initial particle trajectories but different protoplanetary disk parameters. Each run considers 2000 stars with a single planet, and uses 40 PSO particles in each run. The panels show the evolution of the best-fit values for 3 model parameters versus PSO iteration number.}
\end{figure}

Figure~15 shows the evolution of the best-fit values for three of the model parameters in the runs with 40 particles. These three parameters (the migration rate exponent $x$, the maximum gas accretion factor $\fgas$, and the pebble isolation factor $\fiso$) are chosen here since they tend to vary more than the others during the course of a PSO run. As the figure shows, runs with different initial particle positions and velocities follow very different paths, but 9 out of the 10 cases converge to the same solution. The figure also shows that the best fit solution can undergo large jumps, representing a big improvement, after many iterations when one of the particles finds a new global minimum.

Finally, Figure 16 shows 10 PSO runs with identical initial trajectories for the particles but different random parameters for the protoplanetary disks. Each run uses 40 particles. Unlike the previous case, there is no reason the runs should converge to a single solution since they are considering different sets of protoplanetary disks. After following the same path for the first hundred or so iterations, the 10 runs do indeed diverge, and end up finding a variety of solutions. This set of runs demonstrates that the evolution is not strongly affected by the particles' initial trajectories.

\newpage
\startlongtable
\begin{deluxetable}{ccccccc}
\tablecaption{Frequency of planets for Solutions A and B compared with the observed frequencies.}
\tablehead{
\colhead{Constraint 1} & \colhead{Constraint 2} & \colhead{Observed} & \colhead{Solution A} & \colhead{N $\sigma$} & \colhead{Solution B} & 
\colhead{N $\sigma$} \\
}
\startdata
$R>\rearth$     & $5<P<7.3$ d               & $0.0160\pm0.0040$  & 0.0265 & 2.63 & 0.0215 & 1.38 \\
$R>\rearth$     & $7.3<P<10.8$ d          & $0.0270\pm0.0070$  & 0.0410 & 2.00 & 0.0415 & 2.07 \\
$R>\rearth$     & $10.8<P<15.8$ d        & $0.0460\pm0.0100$  & 0.0265 & -1.95 & 0.0395 & -0.65 \\
$R>\rearth$     & $15.8<P<23.2$ d        & $0.0520\pm0.0130$  & 0.0340 & -1.39 & 0.0345 & -1.35 \\
$R>\rearth$     & $23.2<P<34.1$ d        & $0.0500\pm0.0130$  & 0.0255 & -1.89 & 0.0375 & -0.96 \\
$R>\rearth$     & $34.1<P<50$ d           & $0.0580\pm0.0170$  & 0.0335 & -1.44 & 0.0380 & -1.18 \\
$M>50\mearth$ & $P<10$ d                   & $0.0083\pm0.0034$  & 0.0015 & -2.00 & 0         & -2.44 \\
$M>50\mearth$ & $10<P<85$ d             & $0.0164\pm0.0055$  & 0.0095 & -1.26 & 0.0040 & -2.26 \\
$M>50\mearth$ & $85<P<400$ d           & $0.0290\pm0.0072$  & 0.0240 & -0.69 & 0.0235 & -0.76 \\
$M>50\mearth$ & $P<10$ y                   & $0.1390\pm0.0170$  & 0.0780 & -3.59 & 0.1205 & -1.09 \\
$M>\mjup$         & $P<11.5$ d               & $0.0043\pm0.0030$  & 0.0010 & -1.10 & 0         & -1.43 \\
$M>\mjup$         & $P<100$ d                & $0.0085\pm0.0040$  & 0.0110 & 0.63 & 0.0020 & -1.63 \\
$M>\mjup$         & $P<365$ d                & $0.0190\pm0.0060$  & 0.0275 & 1.42 & 0.0150 & -0.67 \\
$M>\mjup$         & $P<1022$ d              & $0.0390\pm0.0090$  & 0.0400  & 0.11 & 0.0365 & -0.28 \\
$M>\mjup$         & $P<1896$ d              & $0.0460\pm0.0100$  & 0.0535  & 0.75 & 0.0530 & 0.70 \\
$M>\mjup$         & $P<4080$ d              & $0.0890\pm0.0140$  & 0.0685 & -1.46 & 0.0975 & 0.60 \\
$1.16<R/\rearth<1.29$ & $P<100$ d      & $0.0780\pm0.0170$  & 0.0230 & -3.24 & 0.0025 & -4.44 \\
$1.29<R/\rearth<1.43$ & $P<100$ d      & $0.0800\pm0.0130$  & 0.0245 & -4.27 & 0.0020 & -6.00 \\
$1.43<R/\rearth<1.59$ & $P<100$ d      & $0.0530\pm0.0110$  & 0.0260 & -2.46 & 0.0025 & -4.59 \\
$1.59<R/\rearth<1.77$ & $P<100$ d      & $0.0334\pm0.0092$  & 0.0180 & -1.67 & 0.0035 & -3.25 \\
$1.77<R/\rearth<1.97$ & $P<100$ d      & $0.0500\pm0.0100$  & 0.0190 & -3.10 & 0.0065 & -4.35 \\
$1.97<R/\rearth<2.19$ & $P<100$ d      & $0.0860\pm0.0160$  & 0.0200 & -4.13 & 0.0040 & -5.13 \\
$2.19<R/\rearth<2.43$ & $P<100$ d      & $0.0980\pm0.0160$  & 0.0150 & -5.19 & 0.0045 & -5.84\\
$2.43<R/\rearth<2.70$ & $P<100$ d      & $0.0770\pm0.0160$  & 0.0125 & -4.03 & 0.0055 & -4.47 \\
$2.70<R/\rearth<3.00$ & $P<100$ d      & $0.0530\pm0.0120$  & 0.0090 & -3.67 & 0.0080 & -3.75 \\
$3.00<R/\rearth<3.33$ & $P<100$ d      & $0.0316\pm0.0089$ & 0.0075 & -2.71 & 0.0060 & -2.88 \\
$3.33<R/\rearth<3.70$ & $P<100$ d      & $0.0242\pm0.0066$ & 0.0050 & -2.91 & 0.0110 & -2.00 \\
$3.70<R/\rearth<4.12$ & $P<100$ d      & $0.0094\pm0.0057$ & 0.0105 &  0.19 & 0.0110 & 0.28 \\
$0.75<R/\rearth<2.5$   & $50<P<300$ d & $0.77\pm0.245$      & 0.0440 & -2.96 & 0.0170 & -3.07 \\
$0.1<R/R_{\rm Jup}<1$ & $2<P<25$ y   & $2.00\pm0.72$         & 0.1490 & -2.57 & 0.1915 & -2.51 \\
$M>13\mjup$               & -                       & $0.0069\pm0.0021$ & 0.0070 & 0.05 & 0.0065 & -0.19 \\
\enddata
\end{deluxetable}

\newpage
\startlongtable
\begin{deluxetable}{ccccccc}
\tablecaption{Frequency of planets for Solutions C and D compared with the observed frequencies.}
\tablehead{
\colhead{Constraint 1} & \colhead{Constraint 2} & \colhead{Observed} & \colhead{Solution C} & \colhead{N $\sigma$} & \colhead{Solution D} & 
\colhead{N $\sigma$} \\
}
\startdata
$R>\rearth$     & $5<P<7.3$ d              & $0.0160\pm0.0040$  & 0.0260 & 2.50 & 0.0175 & 0.38 \\
$R>\rearth$     & $7.3<P<10.8$ d         & $0.0270\pm0.0070$  & 0.0380 & 1.57 & 0.0315 & 0.64 \\
$R>\rearth$     & $10.8<P<15.8$ d       & $0.0460\pm0.0100$  & 0.0290 & -1.70 & 0.0415 & -0.45 \\
$R>\rearth$     & $15.8<P<23.2$ d       & $0.0520\pm0.0130$  & 0.0275 & -1.89 & 0.0485 & -0.27 \\
$R>\rearth$     & $23.2<P<34.1$ d       & $0.0500\pm0.0130$  & 0.0340 & -1.23 & 0.0570 & 0.54 \\
$R>\rearth$     & $34.1<P<50$ d          & $0.0580\pm0.0170$  & 0.0515 & -0.38 & 0.0765 & 1.09 \\
$M>50\mearth$ & $P<10$ d                   & $0.0083\pm0.0034$  & 0.0000 & -2.44 & 0         & -2.44 \\
$M>50\mearth$ & $10<P<85$ d             & $0.0164\pm0.0055$  & 0.0085 & -1.44 & 0         & -2.98 \\
$M>50\mearth$ & $85<P<400$ d           & $0.0290\pm0.0072$  & 0.0245 & -0.63 & 0         & -4.03 \\
$M>50\mearth$ & $P<10$ y                   & $0.1390\pm0.0170$  & 0.0970 & -2.47 & 0         & -8.18 \\
$M>\mjup$         & $P<11.5$ d               & $0.0043\pm0.0030$  & 0.0000 & -1.43 & 0         & -1.43 \\
$M>\mjup$         & $P<100$ d                & $0.0085\pm0.0040$  & 0.0070 & -0.38 & 0          & -2.13 \\
$M>\mjup$         & $P<365$ d                & $0.0190\pm0.0060$  & 0.0255 & 1.08 & 0          & -3.17 \\
$M>\mjup$         & $P<1022$ d              & $0.0390\pm0.0090$  & 0.0440 & 0.56 & 0          & -4.33 \\
$M>\mjup$         & $P<1896$ d              & $0.0460\pm0.0100$  & 0.0610  & 1.50 & 0          & -4.60 \\
$M>\mjup$         & $P<4080$ d              & $0.0890\pm0.0140$  & 0.0875 & -0.11 & 0          & -6.36 \\
$1.16<R/\rearth<1.29$ & $P<100$ d      & $0.0780\pm0.0170$  & 0.0420 & -2.12 & 0.0810 &  0.18 \\
$1.29<R/\rearth<1.43$ & $P<100$ d      & $0.0800\pm0.0130$  & 0.0320 & -3.69 & 0.0555 & -1.89 \\
$1.43<R/\rearth<1.59$ & $P<100$ d      & $0.0530\pm0.0110$  & 0.0250 & -2.55 & 0.0410 & -1.09 \\
$1.59<R/\rearth<1.77$ & $P<100$ d      & $0.0334\pm0.0092$  & 0.0230 & -1.13 & 0.0325 & -0.10 \\
$1.77<R/\rearth<1.97$ & $P<100$ d      & $0.0500\pm0.0100$  & 0.0195 & -3.05 & 0.0350 & -1.50 \\
$1.97<R/\rearth<2.19$ & $P<100$ d      & $0.0860\pm0.0160$  & 0.0175 & -4.28 & 0.0305 & -3.47 \\
$2.19<R/\rearth<2.43$ & $P<100$ d      & $0.0980\pm0.0160$  & 0.0120 & -5.38 & 0.0275 & -4.41\\
$2.43<R/\rearth<2.70$ & $P<100$ d      & $0.0770\pm0.0160$  & 0.0130 & -4.00 & 0.0195 & -3.59 \\
$2.70<R/\rearth<3.00$ & $P<100$ d      & $0.0530\pm0.0120$  & 0.0145 & -3.21 & 0.0145 & -3.21 \\
$3.00<R/\rearth<3.33$ & $P<100$ d      & $0.0316\pm0.0089$ & 0.0090 & -2.54 & 0.0065 & -2.82 \\
$3.33<R/\rearth<3.70$ & $P<100$ d      & $0.0242\pm0.0066$ & 0.0180 & -0.94 & 0.0015 & -3.44 \\
$3.70<R/\rearth<4.12$ & $P<100$ d      & $0.0094\pm0.0057$ & 0.0175 &  1.42 & 0.0010  & -1.47 \\
$0.75<R/\rearth<2.5$   & $50<P<300$ d & $0.77\pm0.245$      & 0.4935 & -1.13 & 0.5235 & -1.00 \\
$0.1<R/R_{\rm Jup}<1$ & $2<P<25$ y   & $2.00\pm0.72$         & 0.5600 & -2.00 & 0.5945 & -1.95 \\
$M>13\mjup$               & -                       & $0.0069\pm0.0021$ & 0.0070 & 0.05  & 0           & -3.29 \\
\enddata
\end{deluxetable}

%
%

\end{document}